\documentclass[useAMS,usenatbib,twocolappendix,apj]{emulateapj}

\usepackage{epsfig}
\usepackage{graphicx}
\usepackage{epstopdf}
\DeclareGraphicsExtensions{.ps}
\usepackage{times}
\usepackage{natbib}
\usepackage{amssymb}
\usepackage{amsmath}

\newif\ifAMStwofonts
\AMStwofontstrue

\newcommand{\be}{\begin{equation}}
  \newcommand{\ee}{\end{equation}}
\newcommand{\ba}{\begin{eqnarray}}
  \newcommand{\ea}{\end{eqnarray}}
\newcommand{\brr}{\begin{array}}
  \newcommand{\err}{\end{array}}
\newcommand{\bc}{\begin{center}}
  \newcommand{\ec}{\end{center}}

\newcommand{\hk}{\,h^{-1}{\rm kpc}}
\newcommand{\msun}{\,{\rm M}_\odot}
\newcommand{\msunh}{\,h_{70}^{-1}{\rm M}_\odot}
\newcommand{\hMpc}{\mbox{$h^{-1}{\rm{Mpc}}~$}}
\newcommand{\rvir}{\mbox{$R_{\rm{vir}}$}}
\newcommand{\mvir}{\mbox{$M_{\rm{vir}}$ }}

\newcommand{\vel}{\,{\rm km\,s^{-1}}}

\newcommand{\mincir}{\raise
  -2.truept\hbox{\rlap{\hbox{$\sim$}}\raise5.truept \hbox{$<$}\ }}
\newcommand{\magcir}{\raise
  -2.truept\hbox{\rlap{\hbox{$\sim$}}\raise5.truept \hbox{$>$}\ }}
\newcommand{\siml}{\raise
  -2.truept\hbox{\rlap{\hbox{$\sim$}}\raise5.truept \hbox{$<$}\ }}
\newcommand{\simg}{\raise
  -2.truept\hbox{\rlap{\hbox{$\sim$}}\raise5.truept \hbox{$>$}\ }}

\def\Munich{1}
\def\ExcellenceCluster{2}
\def\MPE{3}
\def\MPA{4}

\begin{document}

\shorttitle{Cluster Mass Calibration with Velocity Dispersions}
\shortauthors{Saro et al.}

\title{Toward Unbiased Galaxy Cluster Masses from Line of Sight Velocity Dispersions} 

\author{  
Alex Saro\altaffilmark{\Munich,\ExcellenceCluster}, 
Gurvan Bazin\altaffilmark{\Munich,\ExcellenceCluster}, 
Joseph J. Mohr\altaffilmark{\Munich,\ExcellenceCluster,\MPE}, 
and Klaus Dolag\altaffilmark{\Munich,\MPA}}

\altaffiltext{\Munich}{Department of Physics, Ludwig-Maximilians-Universit\"{a}t, Scheinerstr.\ 1, 81679 M\"{u}nchen, Germany}
\altaffiltext{\ExcellenceCluster}{Excellence Cluster Universe, Boltzmannstr.\ 2, 85748 Garching, Germany}
\altaffiltext{\MPE}{Max-Planck-Institut f\"{u}r extraterrestrische Physik, Giessenbachstr.\ 85748 Garching, Germany}
\altaffiltext{\MPA}{Max-Planck-Institut f\"{u}r Astrophysik, Karl-Schwarzschild-Str.\ 1, 85748 Garching, Germany}

\begin{abstract}
We study the use of red sequence selected galaxy spectroscopy for
unbiased estimation of galaxy cluster masses.  We use the publicly 
available galaxy catalog produced using the semi-analytic 
model of \citet{delucia07} on the Millenium Simulation (\citealt{Springel_etal_2005}).
We make mock observations to mimic the selection of the galaxy sample, the
interloper rejection and the dispersion measurements for large numbers
of simulated clusters spanning a wide range in mass and redshift. We
explore the impacts on selection using galaxy color, projected separation
from the cluster center, and galaxy luminosity.  We probe for biases
and characterize sources of scatter in the relationship between
cluster virial mass and velocity dispersion.  
We identify and characterize the following sources of bias and scatter:
  intrinsic properties of halos in the form of halo triaxiality,
  dynamical friction of red luminous galaxies and interlopers. We show
  that due to halo triaxiality the intrinsic scatter of estimated
  line-of-sight dynamical mass is about three times larger ($30-40
  \%$) than the one estimated using the 3D velocity dispersion ($\sim
  12 \%$) and a small bias ($\lesssim 1\%$) is induced. Furthermore we
  find evidence of increasing scatter as a function of redshift and
  provide a fitting formula to account for it. We characterize the
  amount of bias and scatter introduced by dynamical friction when
  using subsamples of red-luminous galaxies to estimate the velocity
  dispersion. We study the presence of interlopers in spectroscopic
  samples and their effect on the estimated cluster dynamical mass.
Our results show that while cluster velocity dispersions extracted
from a few dozen red sequence selected galaxies do not provide precise
masses on a single cluster basis, an ensemble of cluster velocity
dispersions can be combined to produce a precise calibration of a
cluster survey mass--observable relation.  Currently, disagreements in the
literature on simulated subhalo velocity dispersion- mass relations place a systematic
floor on velocity dispersion mass calibration at the 15\% level in mass.  We show that
the selection related uncertainties are small by comparison, providing hope that
with further improvements to numerical studies this systematic floor can be substantially reduced.

\end{abstract}

%\begin{keywords}
%  Cosmology: theory -- galaxies: clusters -- methods: N-body
%  simulations, numerical --  hydrodynamics
%\end{keywords}

%%%%%%%%%%%%%%%%%%%%%%%%%%%%%%%%%%%%%%%%%%%%%%%%%%%%%%%%%%%%%%%%%%%%%%%%%%%%%%%
\section{Introduction} 
\label{sec:intro} 

Clusters of galaxies are the most massive collapsed objects in the
Universe and sit at the top of the hierarchy of non--linear
structures.  They were first identified as over--dense regions in the
projected number counts of galaxies (e.g. \citealt{Abell58},
\citealt{ZW68.1}).  However nowadays clusters can be identified over
the whole electro-magnetic range, including as X-ray sources
(e.g. \citealt{boehringer00}, \citealt{pacaud07},
\citealt{vikhlinin09b}, \citealt{suhada12}), as optically
overdensities of red galaxies ( \citealt{Gladders05},
\citealt{Koester07}, \citealt{Hao10}, \citealt{Szabo11}) and as
distorsions of the cosmic microwave background as predicted by
\citet{Sunyaev72} (e.g. \citealt{vanderlinde10}, \citealt{Marriage11},
\citealt{Planck11}).

Given the continuous improvement in both spatial and spectral
resolution power of modern X--ray, optical and infrared telescopes,
more and more details on the inner properties of galaxy clusters have
been unveiled in the last decade. These objects, that in a first
approximation were thought to be virialized and spherically symmetric,
have very complex dynamical features -- such as strong asymmetries and
clumpiness (e.g. \citealt{geller82}, \citealt{dressler88},
\citealt{mohr95}) -- witnessing for violent processes being acting or
having just played a role.  They exhibit luminosity and temperature
functions which are not trivially related to their mass function, as
one would expect for virialized gravitation--driven objects. Moreover,
the radial structure of baryons' properties is far from being
completely understood: a number of observational facts pose a real
challenge to our ability in modeling the physics of the intracluster
medium and the closely related physics of the galaxy
population. Indeed a number of different physical processes are acting
together during the formation and evolution of galaxy clusters. Gas
cooling, star formation, chemical enrichment, feedback from supernovae
explosions and from active galactic nuclei, etc are physical processes
at the base of galaxy formation, which are difficult to disentangle
(e.g. see \citealt{benson10} for a recent review on galaxy
  formation models).

Line-of-sight galaxy velocities in principle provide a measure of the
depth of the gravitational potential well and therefore can be used to
estimate cluster masses. Furthermore, galaxy dynamics are expected to
be less affected by the complex baryonic physics affecting the intra
cluster medium. Thus, one would naively expect a mass function defined on the
basis of velocity dispersion to be a good proxy of the underlying
cluster mass. However a number of possible systematics can affect
dynamical mass estimation and must be carefully take into
account. \citet{biviano06} for example studied a sample of 62 clusters
at redshift $z = 0$ from a $\Lambda CDM$ cosmological hydrodynamical
simulation. They estimated virial masses from both dark matter (DM)
particles and simulated galaxies in two independent ways: a virial
mass estimator corrected for the surface pressure term, and a mass
estimator based entirely on the velocity dispersion $\sigma_v$. They
also modeled interlopers by selecting galaxies within cylinders of
different radius and length $192 \hMpc$ and applying interloper
removal techniques. They found that the mass estimator based entirely
on velocity dispersions is less sensitive on the radially dependent
incompleteness. Furthermore the effect of interlopers is smaller if
early type galaxies, defined in the simulations using their mean
redshift of formation, are selected. However, the velocity dispersion
of early type galaxies is biased low with respect to DM particles.
\citet{evrard08} analysed a set of different simulations with
different cosmologies, physics and resolutions and found that the 3D
velocity dispersion of DM particles within the virial radius can be
expressed as a tight function of the halo virial
mass\footnote{Throughout the text, we will refer to \mvir as the mass
  contained within a radius \rvir encompassing a mean density equal to
  200 $\rho_{c}$, where $\rho_{c}$ is the critical cosmic density.},
regardless of the simulation details. They also found the scatter
about the mean relation is nearly log-normal with a low standard
deviation $\sigma_{ln \sigma} \simeq 0.04$. In a more recent work,
\citet{white10} used high resolution N-body simulations to study how
the influence of large scale structure could affect different physical
probes, including the velocity dispersion based upon sub-halo
dynamics. They found that the highly anisotropic nature of infall
material into clusters of galaxies and their intrinsic triaxiality is
responsible for the large variance of the 1D velocity dispersion under
different lines of sight. They also studied how different interloper
removal techniques affect the velocity dispersion and the stability of
velocity dispersion as a function of the number of sub-halos used to
estimate it. They found that only when using small numbers of
sub-halos ($\lesssim 30$) is the line of sight velocity dispersion
biased low and the scatter significantly increases with respect to the
DM velocity dispersion. Furthermore the effect of interlopers is
different for different interloper rejection techniques and can
significantly increase the scatter and bias low velocity dispersion
estimates.

Currently IR, SZE and X-ray cluster surveys are delivering significant
numbers of clusters at redshifts $z>1$ (e.g. \citealt{stanford05},
\citealt{staniszewski09}, \citealt{fassbender11},
\citealt{williamson11}, \citealt{reichardt12}).  Mass calibration of
these cluster samples is challenging using weak lensing, making
velocity dispersion mass estimates particularly valuable.  At these
redshifts it is also prohibitively expensive to obtain spectroscopy of
large samples of cluster galaxies, and therefore dispersion
measurements must rely on small samples of 20 to 30 cluster members.
This makes it critically important to understand how one can best use
the dynamical information of a few dozen of the most luminous cluster
galaxies to constrain the cluster mass.  It is clear that with such a
small sample one cannot obtain precise mass estimates of individual
clusters.  However, for mass calibration of a cluster SZE survey, for
example, an {\it unbiased} mass estimator with a large statistical
uncertainty is still valuable.

In this work we focus on the characterisation of dynamical mass of
clusters with particular emphasis on high-z clusters with a small number of measured galaxy velocities.  The plan of the paper is as follows. In Sec. \ref{sec:sims} we briefly
introduce the simulation describe the adopted semi-analytic model, and
in Sec. \ref{sec:Res} we present the results of our analysis. Finally,
in Sec. \ref{sec:Concl}, we summarise our findings and give our
conclusions.

%%%%%%%%%%%%%%%%%%%%%%%%%%%%%%%%%%%%%%%%%%%%%%%%%%%%%%%%%%%%%%%%%%%%%%%%%%%%%%%
\begin{table} 
  \centering
  \caption{The redshift--number distribution of the 22,484 clusters 
    with \mvir$>10^{14} \msun$
    analysed in this work at different redshift. Column 1: redshift
    $z$; Column 2: number of clusters $N_{clus}$.}
  \begin{tabular}{ll}
    $z$ & $N_{clus}$ \\  
    \hline 
    0.00 & 3133\\ 
    0.09 & 2953\\ 
    0.21 & 2678\\
    0.32 & 2408\\
    0.41 & 2180\\
    0.51 & 1912\\
    0.62 & 1635\\
    0.75 & 1292\\
    0.83 & 1152\\
    0.91 & 1020\\
    0.99 & 867\\
    1.08 & 702\\
    1.17 & 552\\
    
  \end{tabular}
  \label{t:clus}
\end{table}

\begin{figure*}
  \centerline{ \hbox{ \psfig{file=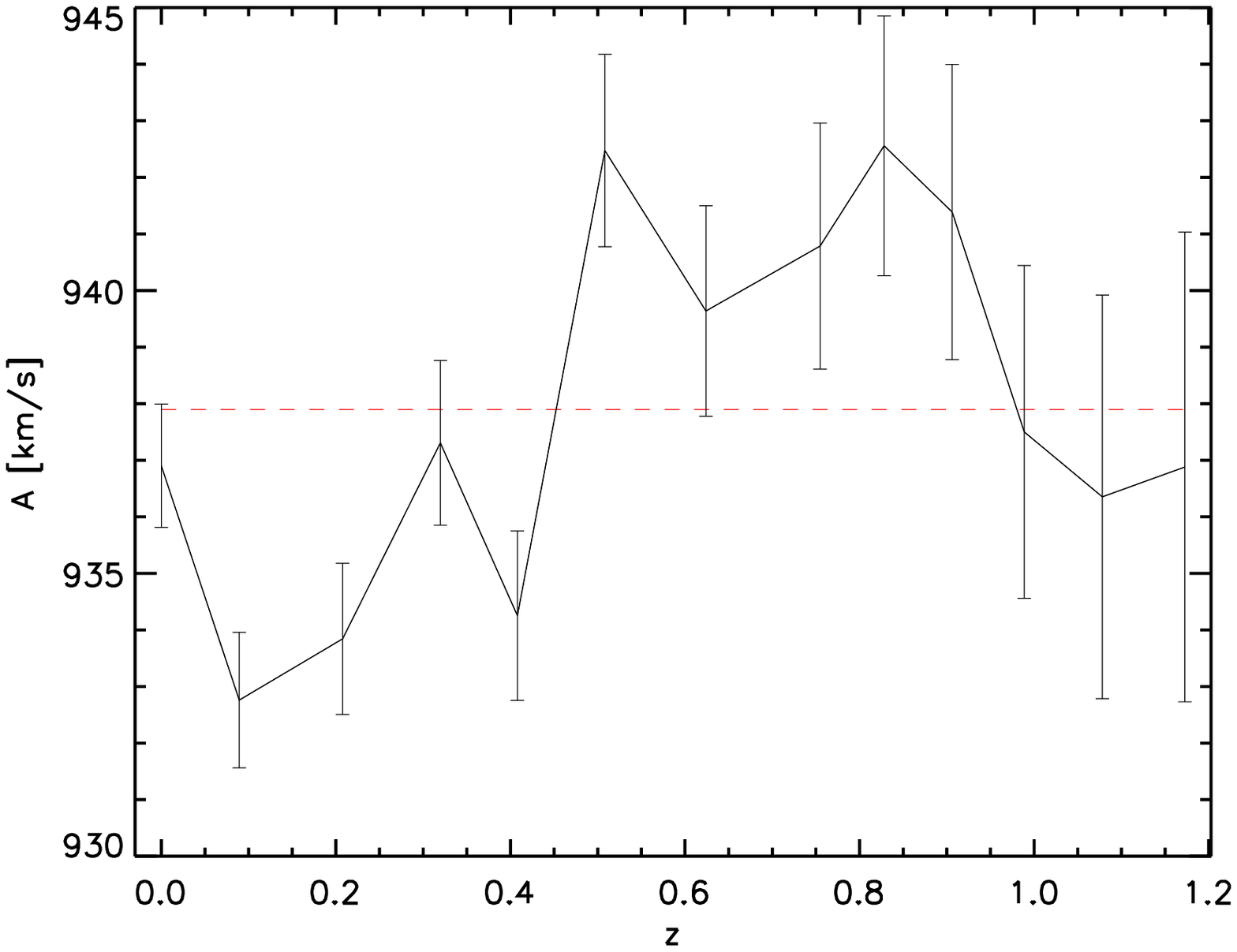,width=9.0cm}
      \psfig{file=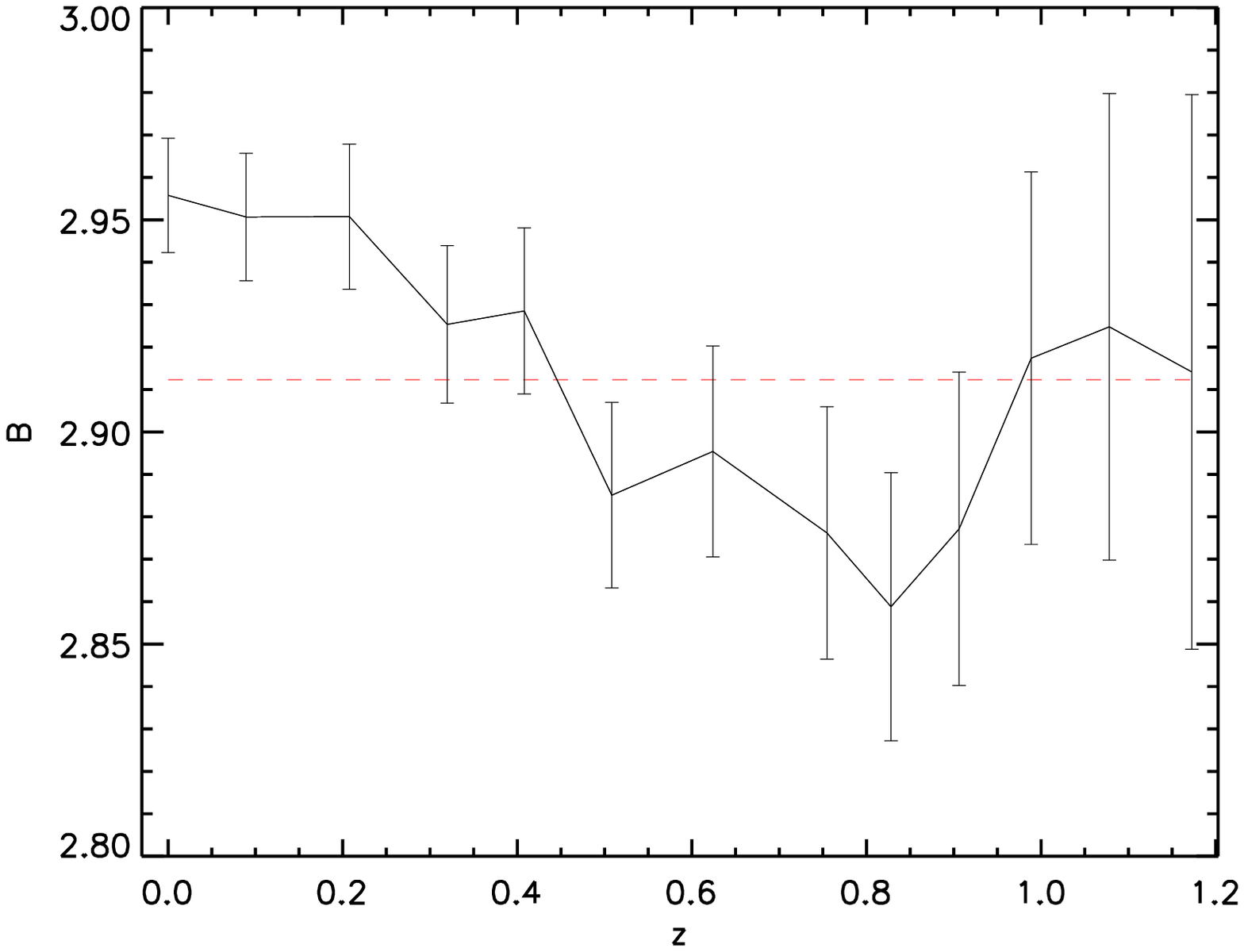,width=9.0cm}
}}
  \caption{The evolution of the normalization $A$ (left panel) and
    slope $B$ (right panel) parameters used to fit the relation
    between the 3D velocity dispersion of all the galaxies within
    \rvir and the virial mass of each cluster (eq. \ref{eq:fit}). Red
    horizontal dashed lines represent the mean value.  The slope is
    moderately shallower than the self-similar expectation.
  }
  \label{fi:EV}
\end{figure*}

\section{Input Simulation} 
\label{sec:sims}
 
This analysis is based on the publicly available galaxy catalogue
produced using the semi-analytic model (SAM) by \citet{delucia07} on
the Millennium Simulation (\citealt{Springel_etal_2005}).  The
Millennium Simulation adopts the following values for the parameters
of a flat $\Lambda $ cold dark matter model: $\Omega_{DM} = 0.205$ and
$\Omega_b = 0.045$ for the densities in cold dark matter and baryons
at redshift $z = 0$, $\sigma_8 = 0.9$ for the rms linear mass
fluctuation in a sphere of radius $8 \hMpc$, $h = 0.73$ for the
present dimensionless value of the Hubble constant and $n = 1$ for the
spectral index of the primordial fluctuation. The simulation follows
the evolution of $2160^3$ dark matter particles from $z = 127$ to the
present day within a cubic box of $500 \hMpc$ on a side. The
individual dark matter particle mass is $8.6 \times 10^8 h^{-1}
\msun$.  The simulation was carried out with the massively parallel
GADGET-2 code (\citealt{Springel05}). Gravitational forces were
computed with the TreePM method, where long-range forces are
calculated with a classical particle-mesh method while short-range
forces are determined with a hierarchical tree approach
(\citealt{barnes86}). The gravitational force has a Plummer-equivalent
comoving softening of $5 \hk$, which can be taken as the spatial
resolution of the simulation.  Full data are stored 64 times spaced
approximately equally in the logarithm of the expansion factor. Dark
matter halos and subhalos were identified with the friends-of-friends
(FOF; \citealt{davis85})) and SUBFIND (\citealt{springel01})
algorithms, respectively. Based on the halos and subhalos within all
the simulation outputs, detailed merger history trees were
constructed, which form the basic input required by subsequently
applied semi-analytic models of galaxy formation.

We recall that the SAM we employ builds upon the methodology
originally introduced by \citet{kauffmann99}, \citet{springel01a} and
\citet{delucia04a}. We refer to the original papers for details.

The SAM adopted in this study includes explicitly DM
substructures. This means that the halos within which galaxies form
are still followed even when accreted onto larger systems. As
explained in Springel et al. (2001) and De Lucia et al. (2004), the
adoption of this particular scheme leads to the definition of
different galaxy types. Each FOF group hosts a Central galaxy. This
galaxy is located at the position of the most bound particle of the
main halo, and it is the only galaxy fed by radiative cooling from the
surrounding hot halo medium.  Besides central galaxies, all galaxies
attached to DM substructures are considered as satellite
galaxies. These galaxies were previously central galaxies of a halo
that merged to form the larger system in which they currently
reside. The positions and velocities of these galaxies are followed by
tracing the surviving core of the parent halo. The hot reservoir
originally associated with the galaxy is assumed to be kinematically
stripped at the time of accretion and is added to the hot component of
the new main halo. Tidal truncation and stripping rapidly reduce the
mass of DM substructures (but not the associated stellar mass) below
the resolution limit of the simulation (\citealt{delucia04b}; Gao et
al. 2004). When this happens, we estimate a residual surviving time
for the satellite galaxies using the classical dynamical friction
formula, and we follow the positions and velocities of the galaxies by
tracing the most bound particles of the destroyed substructures.

%%%%%%%%%%%%%%%%%%%%%%%%%%%%%%%%%%%%%%%%%%%%%%%%%%%%%%%%%%%%%%%%%%%%%%%%%%%%%%%
\section{Properties of the Full Galaxy Population}
\label{sec:Res}

\subsection{Intrinsic galaxy velocity dispersion}
\label{sec:intrinsic}
\citet{evrard08} showed that massive dark matter halos adhere to a
virial scaling relation when one expresses the velocity dispersion of
the DM particles as a function of the virial mass of the halo in the
form: \be \sigma_{DM}(M_{vir},z) = \sigma_{DM,15}\left(
\frac{h(z)M_{vir}}{10^{15}\msun}\right)^{\alpha}, \ee where
$\sigma_{DM,15} = 1082.9 \pm 4.0 \vel$ is the typical 3D velocity
dispersion of the DM particles within \rvir\ for a $10^{15}
h^{-1}\msun$ cluster at $z = 0$ and $\alpha = 0.3361 \pm
0.0026$. Similarly, we first compute for each cluster the 3D velocity
dispersion $\sigma_{3D}$ (divided by $\sqrt 3$) of all the galaxies
within \rvir and then fit the relation between $\sigma_{3D}$ and \mvir
in the form of $log(\sigma_{3D}) \propto log(h_{70}(z)\mvir /
10^{15}\msun)$ individually at any of the redshift listed in
Table~\ref{t:clus}. As a result we can express the dynamical mass
$M_{dyn}$ as: \be M_{dyn} = \left( \frac{\sigma_{v}}{A}\right) ^{B}
h_{70}(z)^{-1} 10^{15} \msun,
\label{eq:fit}
\ee where the resulting best fitting values of A and B with their
associated error-bars are shown in Fig. \ref{fi:EV} as a function of
redshift. Dashed horizontal red lines show the average values
which are respectively $\bar A = 938 \pm 3 \vel$ and $\bar B = 2.91 \pm 0.03$.

\begin{figure}
\vspace{-0.3truecm}
\centerline{ \hbox{ \psfig{file=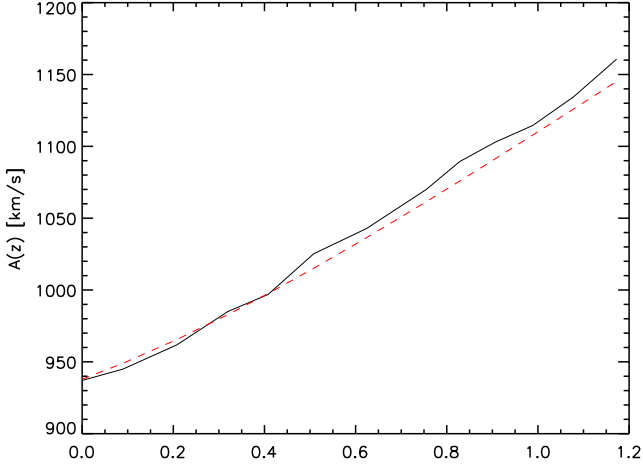,width=9.0cm} }}
\vspace{-0.5truecm}
  \caption{The evolution of the normalisation parameter $A(z)$ of
    Eq. \ref{eq:fit} when no self-similar evolution is taken into account
    (solid black line). Red dashed line is showing the best fitting
    parameter $\bar A \times \left( H(z)/H_0\right) ^{1\over3}$ }
  \label{fi:EVZ}
\end{figure}

After accounting for the differences in the Hubble parameter, our measured normalization
of the galaxy velocity dispersion-- mass relation is within $\lesssim$3\%
of \citet{evrard08}.  This reflects the differences
between the subhalo and DM particle dynamics. As has been previously 
pointed out (e.g. \citealt{gao04}, \citealt{goto05},\citealt{faltenbacher06}, 
\citealt{evrard08}, \citealt{white10}), the velocity bias between galaxies and DM is expected
 to be small $b_v \lesssim 5\%$.  But to be absolutely clear, we
 adopt our measured galaxy velocity dispersion-- mass calibration in the
 analyses that follow.
To better visualize the relative importance of the cosmological
redshift dependence we show in Fig.~\ref{fi:EVZ} the redshift
evolution of the normalisation parameter $A$ (solid black line )when
the fit is made on the relation $log(\sigma_{3D}) \propto
log(h_{70}(z=0) \mvir /10^{15} \msun )$. The expected self-similar
evolution given by $A(z) = \bar A \times E(z)^{1\over3}$ is
highlighted (dashed red line), where the term $\bar A$ is equal to
mean value $\bar A = 938 \vel$ and $E(z)$ describes the universal
expansion history $H(z) = H_0 E(z)$. In other words, Fig.~\ref{fi:EVZ}
shows the typical galaxy velocity dispersion in $\vel$ for a cluster
with \mvir $= 10^{15} \msunh$ as a function of redshift and
demonstrates the nearly self-similar evolution (within $\sim$1\%) 
over the redshift range tested in this work.

\begin{figure*}
  \centerline{ \hbox{ \psfig{file=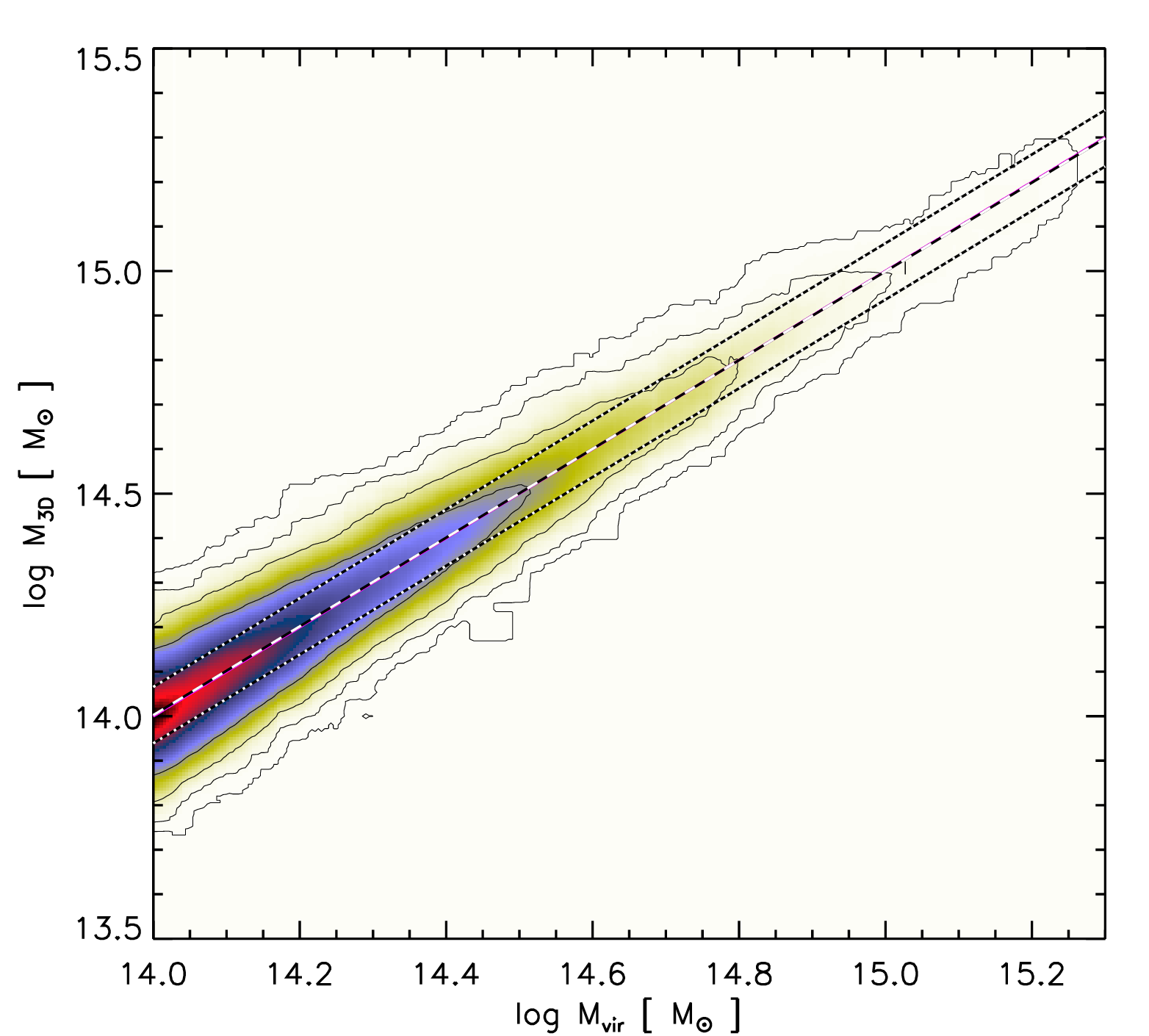,width=9.0cm}
      \psfig{file=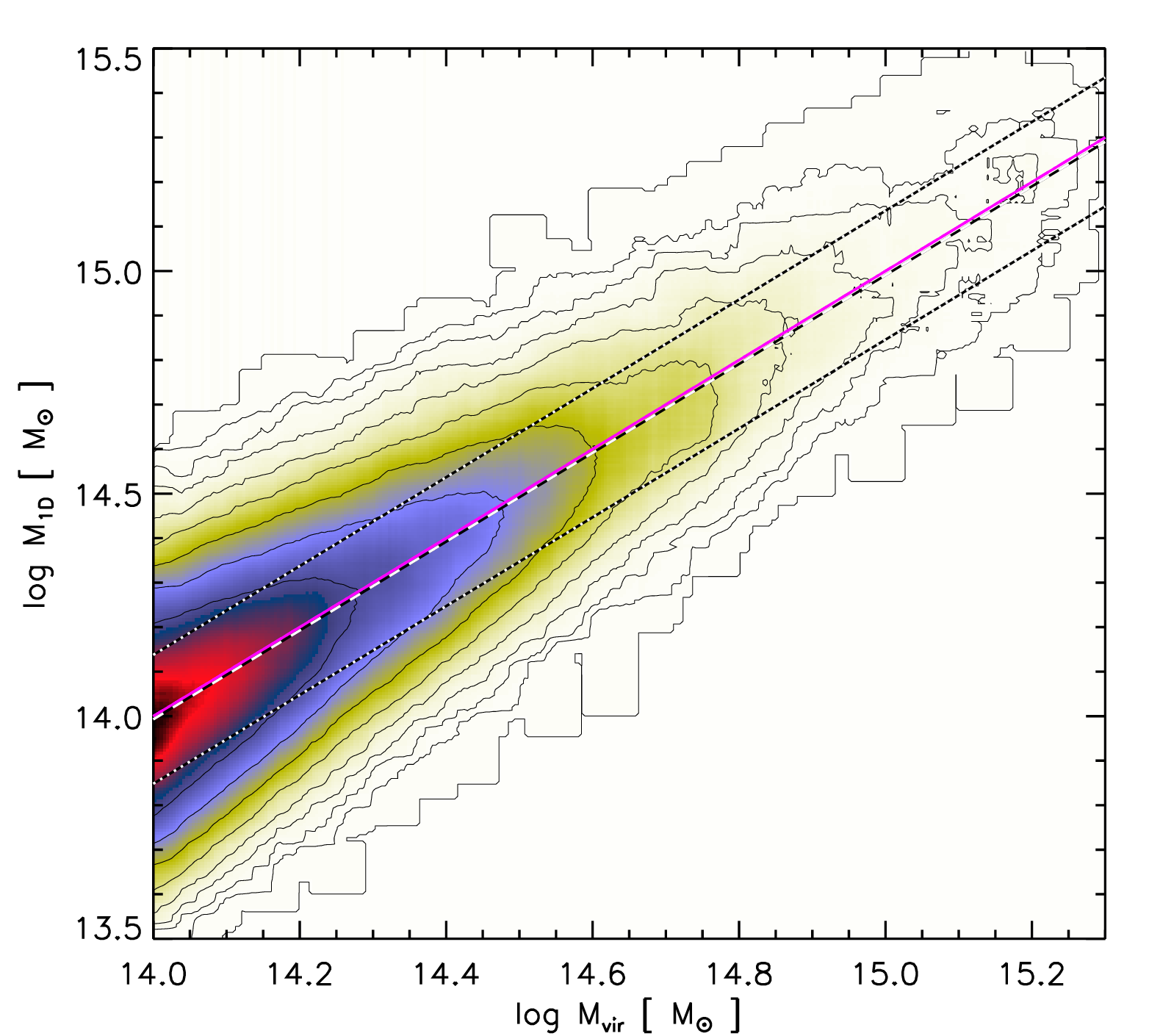,width=9.0cm} }}
  \caption{ The relation between \mvir and the dynamical mass for all
    the clusters in the sample. For each cluster the dynamical mass is
    inferred by applying Eq. \ref{eq:fit} to the 3D velocity
    dispersion divided by $\sqrt{3}$ (left panel) and for each of the
    three projected 1D velocity dispersions (right panel) of all the
    galaxies extracted from the \citet{delucia07} database within
    \rvir\ from the centre of the cluster. The dashed (dotted)
    white-black line is the best fit of the relation (plus and minus
    one $\sigma$) and is virtually indistinguishable from the
    one-to-one relation (dotted-dashed purple line).}
  \label{fi:maps3}
\end{figure*}

For the full sample of clusters analysed (see Table~\ref{t:clus}), we
then compute the dynamical masses by applying Eq.~\ref{eq:fit} to (1)
the 3D galaxy velocity dispersion (divided by $\sqrt 3$) and (2) to
each orthogonal projected 1D velocity dispersion. Fig.~\ref{fi:maps3}
shows the comparison between the virial masses \mvir and the resulting
dynamical masses $M_{3D}$ (left panel) and $M_{1D}$ (right panel) for
the full sample of clusters.  The best fit of the relation (dashed
black and white lines) is virtually indistinguishable from the
one-to-one relation (dotted-dashed purple line) in the case of the 3D
velocity dispersion. On the other hand, in the case of the 1D velocity
dispersion there is a small but detectable difference between the
one-to-one relation and the best fit.  The best fit of the dynamical
mass for the 1D velocity dispersion is about $\lesssim 1\%$ lower than
the one-to-one relation. We will show in Section \ref{sec:triax} that
this difference can be explained in terms of triaxial properties
of halos.  Typical logarithmic scatter of $\sigma_{M_{3D} / M_{vir}}
\simeq 0.145$ and $\sigma_{M_{1D} / M_{vir}} \simeq 0.334$ are
highlighted with dotted black and white lines in $log_{10}$ scale.  We
find that, similar to results by \citet{white10}, using the 1D
velocity dispersion rather than the 3D velocity dispersion increases
the intrinsic log scatter around the mean relation by a factor of
$\sim2.3$.

\begin{figure}
  \centerline{ \hbox{ \psfig{file=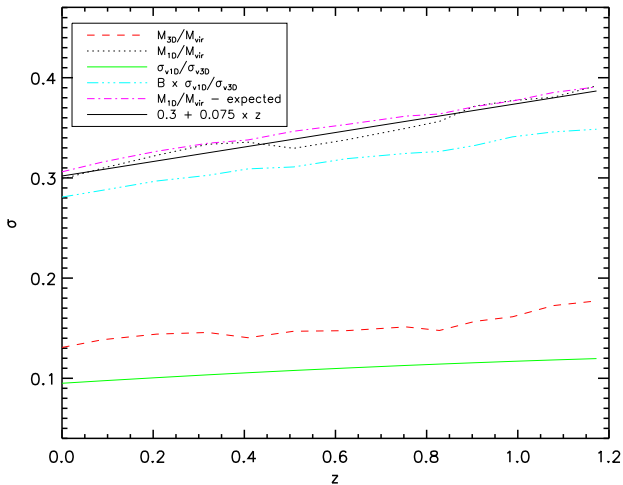,width=9.0cm}}}
  \caption{ The redshift evolution of the logarithmic 1$\sigma$
    scatter for the following quantities: (1) the 3D galaxy velocity
    dispersion mass estimate scatter (dashed red), (2) the 1D galaxy
    velocity dispersion mass estimate scatter (dotted black), (3) a
    fit to \#2 (solid black; Eqn.~\ref{eq:scatfit}), (4) the scatter
    of the 1D velocity dispersion about the 3D dispersion (solid
    green), (5) the same quantity turned into mass scatter using
    Eqn.~\ref{eq:fit} (dashed-dotted blue) and (6) the expected 1D
    dispersion mass scatter (\#2) obtained by quadrature addition of
    \#1 and \#5, as explained in Sec. \ref{sec:intrinsic}
    (dotted-dashed purple; Eqn.~\ref{eq:sumscat}).}
  \label{fi:stddev}
\end{figure}

We further investigate the intrinsic scatter in the relation between the true virial masses and
the dynamical mass estimates in Fig. \ref{fi:stddev}.  Taking $\sigma$ to be the standard deviation of the logarithm of the ratio between the dynamical mass estimate and the virial masses, we show that
in the case of the  3D velocity dispersion (dashed red line) and the 1D
velocity dispersion (dotted black line) the scatter increases with redshift.  
The solid black line shows a linear fit to the evolution of the intrinsic $M_{dyn - 1D}$ scatter and can be expressed as: 
\be 
\sigma_{ln(M_{1D}/M_{vir})} \simeq 0.3 + 0.075 \times z .
\label{eq:scatfit}
\ee
Velocity dispersions are $\sim$25\% less accurate for estimating single cluster masses at $z=1$ than at low redshift.

The logarithmic scatter of the 1D velocity dispersion mass estimator
$\sigma_{M_{1D}}$ around the true mass arises from two sources of
scatter: (1) the logarithmic scatter between the 3D velocity
dispersion mass estimator and the true mass - $\sigma[M_{3D}/M_{vir}]$
(red dashed line in Fig~\ref{fi:stddev}) and (2) the logarithmic
scatter between the 1D and 3D velocity dispersions
$\sigma[\sigma_{1D}/\sigma_{3D}]$ (solid green line).  The expected 1D
dispersion mass scatter is then the quadrature addition of these two
sources: \be \sigma^2_{M_{1D}} \sim \sigma^2[M_{3D}/M_{vir}] + \{\bar
B\times \sigma[\sigma_{1D}/\sigma_{3D}]\}^2,
\label{eq:sumscat}
\ee 
where $\bar B$ is the best fitting slope parameter from
Eq.~\ref{eq:fit}. The expected $\sigma_{M_{1D}}$ estimate from Eqn.~\ref{eq:sumscat} appears as a dotted-dashed purple line in Fig~\ref{fi:stddev}; note that this estimate is in excellent agreement with the directly measured scatter  (dotted black line).  Therefore, we show-- as pointed out by \citet{white10}-- that the dominant contributor to the scatter is the intrinsic triaxial structure of halos. Furthermore its evolution with redshift is also the dominant source of the
increasing scatter of the 1D dynamical mass estimates with redshift.  By comparison, the scatter between the 3D velocity dispersion mass estimator and the true mass $\sigma[ln(M_{3D}/M_{vir})]$, which is reflecting departures from dynamical equilibrium due to ongoing merging in the cluster population, is relatively minor.  Ultimately it is the lack of observational access to the full 3D dynamics and distribution of the galaxies that limits us from precise single cluster dynamical mass estimates.
 
\subsection {Triaxiality}
\label{sec:triax}
\begin{figure*}
  \centerline{\hbox{ \psfig{file=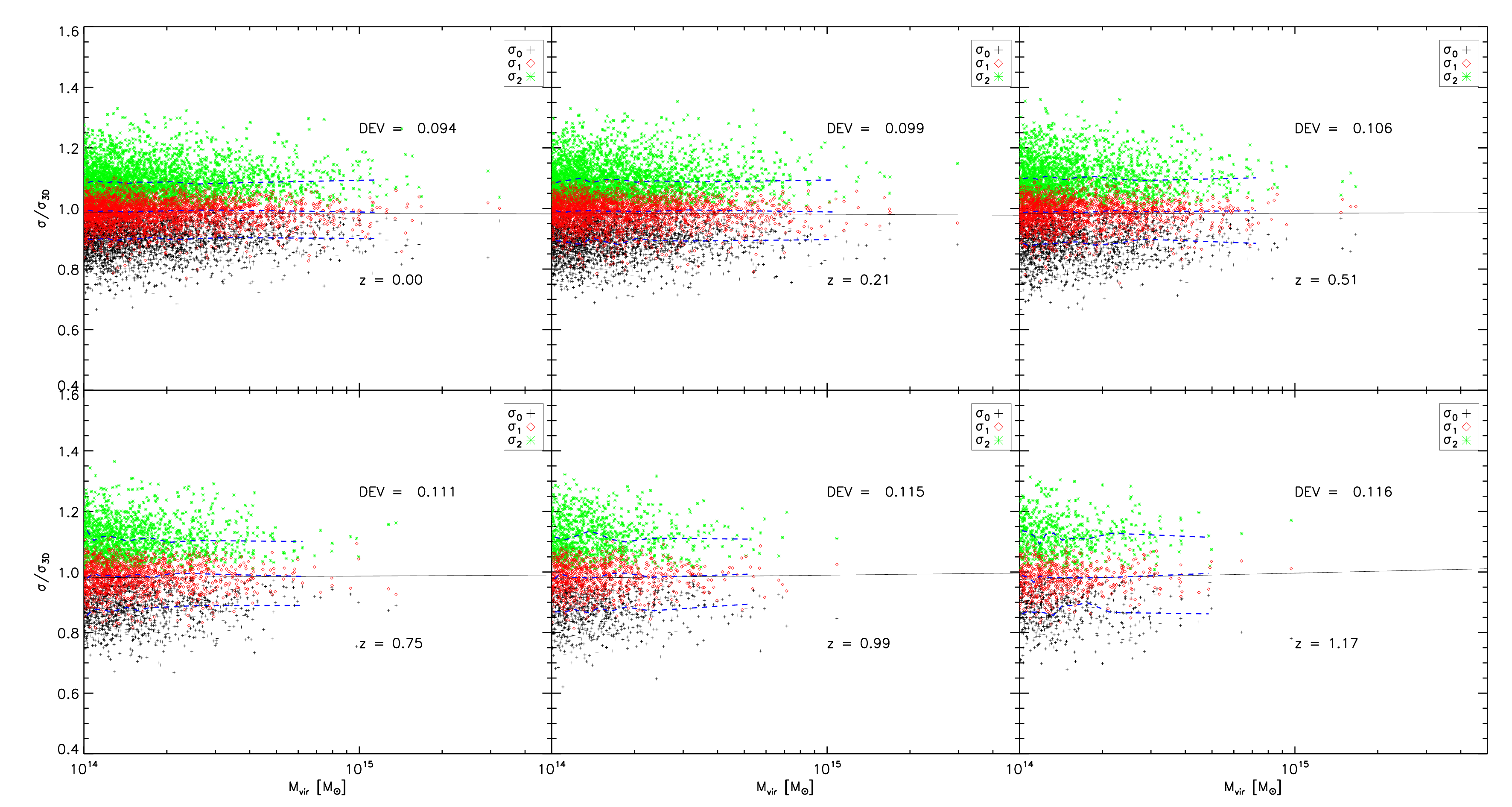,width=20cm} }}
  \caption{We show for each cluster the velocity dispersion along the
    tree major axes of the inertia momentum (black crosses for the
    smaller, red diamonds for the intermediate and green stars for the
    larger) normalized to the 3D velocity dispersion divided by $\sqrt
    3$ as a function of the cluster mass in different redshift
    bins. The black solid line is the best fit of the intermediate axis
    velocity dispersion, and the dashed blue lines are the median and the 16
    and 84 percentile of the full distribution.  DEV is the
    associated standard deviation. }
  \label{fi:prolatez} 
\end{figure*}

The presence of pronounced departures from sphericity in dark matter
halos (\citealt{thomas92}, \citealt{warren92}, \citealt{jing02}), if
not approximately balanced between prolate and oblate systems, could
in principle not only increase the scatter in dynamical mass
estimates, but also lead to a bias. If, for example, clusters were
mainly prolate systems, with one major axis associated to a higher
velocity dispersion and two minor axes with a lower velocity
dispersion, there should be two lines of sight over three associated
with a lower velocity dispersion. This could potentially lead to a
bias in the 1D velocity dispersion with respect to the 3D velocity
dispersion. To quantify this possible bias, we compute the moment of
inertia for each cluster in the sample, and we then calculate the
velocity dispersions along each of the three major axes. As has
been pointed out before (\citealt{tormen97}, \citealt{kasun05}, 
\citealt{white10}) the inertia and velocity tensor are quite well aligned, with typical
misalignment angle of less than $30^\circ$.  In
Fig.~\ref{fi:prolatez}, at each redshift we show the lowest velocity
dispersion $\sigma_0$ with black crosses, the highest $\sigma_2$ with
green stars and the intermediate one $\sigma_1$ with red diamonds
normalized to the 3D velocity dispersion $\sigma_{3D}$ (divided by
$\sqrt 3$). Dashed blue lines are the 16, 50 and 84 percentile of the
full distribution and DEV is the associated standard deviation which,
as expected from Fig.~\ref{fi:stddev} is increasing with redshift. A
perfectly spherical cluster in this plot will therefore appear with
the three points lying all at the value 1, whereas prolate and oblate
systems will have the intermediate velocity dispersions $\sigma_1$
closer to the lower one $\sigma_0$ and to the higher one $\sigma_2$,
respectively. The black solid line is the best fit of the distribution
of the intermediate $\sigma_1$ velocity dispersions and it is very
close to unity, showing that dynamically, clusters do not have a very
strong preference among prolate and oblate systems. Furthermore this
result is true for the range of redshifts and masses we examine here.
  
\begin{figure*}
\centerline{\hbox{
      \psfig{file=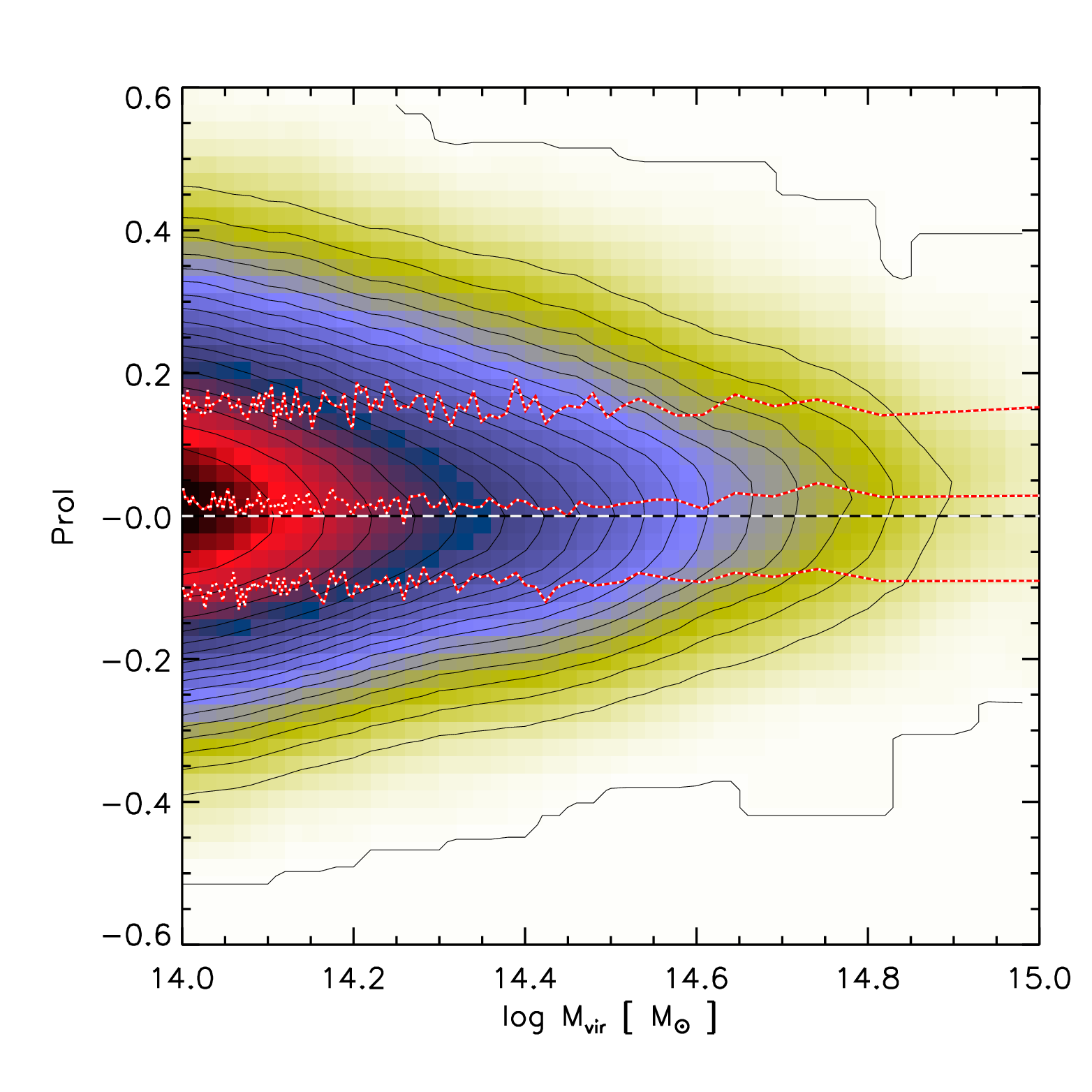,width=9.cm} 
      \psfig{file=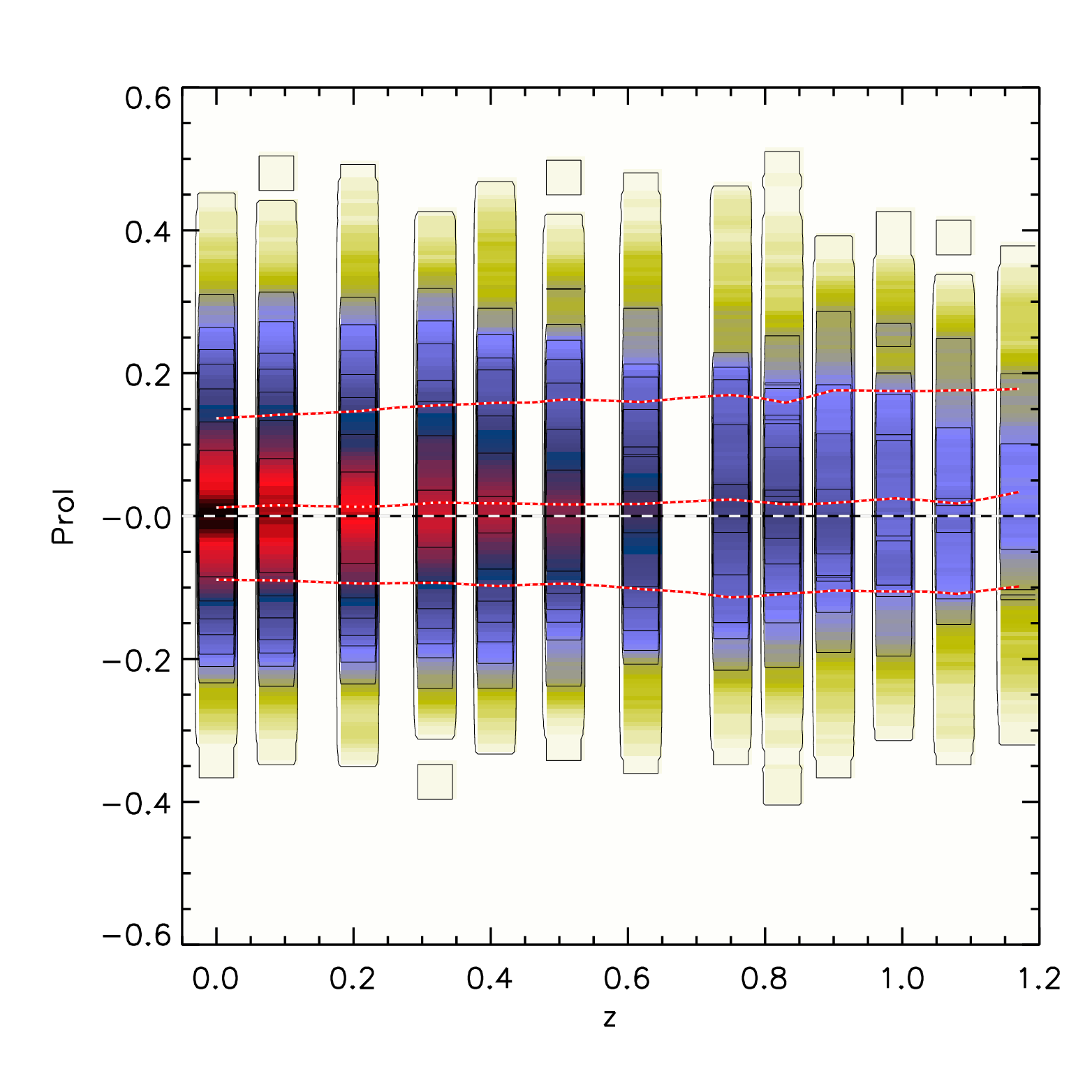,width=9.cm} 
      }}
  \caption
{ The distribution of the prolateness variable $Prol$ (see
  Eqn.~\ref{eq:prol}) as a function of the cluster mass for all the
  clusters at different redshift stacked together (left panel) and as
  a function of redshift (right panel). To guide the eye the dashed
  black and white line is highlighting the value of $Prol = 0$, while
  the dotted red-white lines are respectively the 16, 50 and 84
  percentiles for the two different distributions.  The cluster
  ensemble exhibits a slight preference for prolateness at all masses
  and redshifts.}
  \label{fi:prolate}
\end{figure*}

This can be better seen in Fig.~\ref{fi:prolate}, where we show that
we measure only a mild excess (at $\lesssim 5\%$ level) of prolate
systems. In addition, for each cluster in the sample, we compute a
"prolateness" quantity $Prol$ as: \be Prol = \frac{\left(\sigma_2 -
  \sigma_1\right) - \left(\sigma_1 - \sigma_0\right)}{\sigma_{3D}}.
\label{eq:prol}
\ee A prolate system will have a positive $Prol$ value whereas an
oblate one will have a negative $Prol$.  Fig.~\ref{fi:prolate} shows a
map representing the distribution of the $Prol$ variable as a function
of the cluster mass (left panel) and redshift (right panel).  To
compute the former we stack clusters from all the redshifts, and for
the latter we stack clusters from all masses.  As it is shown in
Fig.~\ref{fi:prolate}, there are no clear dependencies of the $Prol$
variable on the cluster mass or redshift. The slight excess of prolate
over oblate systems at all masses and redshifts would translate into
1D dynamical masses slightly biased towards smaller masses.
Indeed, this is seen as a $\sim1$\% effect in Fig.~\ref{fi:maps3}.

%%%%%%%%%%%%%%%%%%%%%%%%%%%%%%%%%%%%%%%%%%%%%%%%%%%%%%%%%%%%%%%%%%%%%%%%%%%%%%%
\section{Properties of Spectroscopic Subsamples}
\label{sec:selection}

Results in the previous section relied on the full galaxy sample
within each cluster virial region.  We now study the possible
systematics affecting the cluster velocity dispersion and associated
dynamical mass estimates when more realistic selection for the member
galaxies are taken into account.

\begin{figure*}
  \centerline{
    \hbox{
      \psfig{file=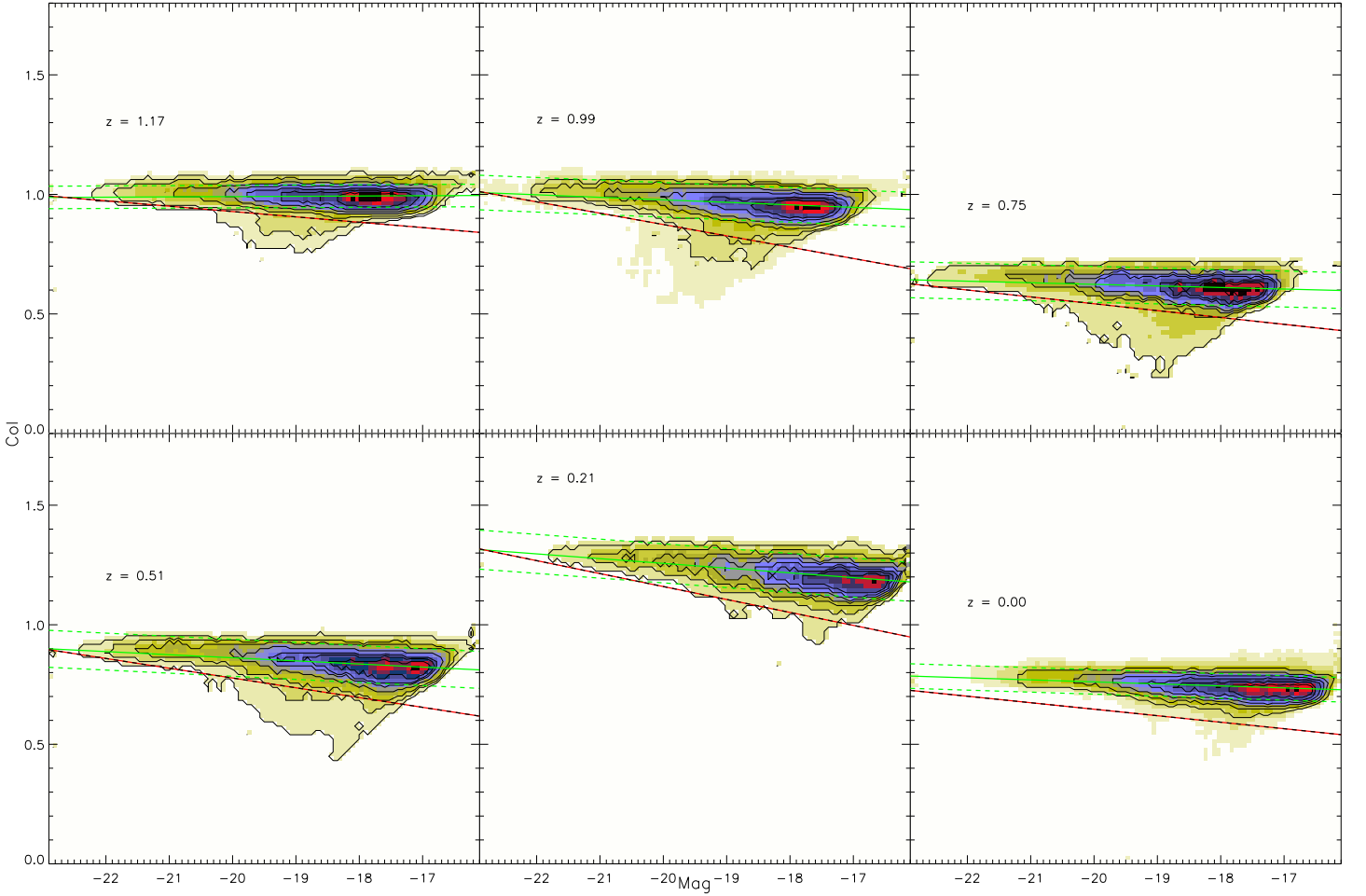,width=20cm}%, height=4cm} 
      }}
    \caption{Color magnitude relation for all the galaxies within
      \rvir at six different redshifts. Color-magnitude relations are
      expressed as $g-r$ as a function of $g$ for redshift z$\leq
      0.35$, $r-i$ as a function of $i$ for redshifts $0.35<$z$\leq
      0.75$ and as $i-z$ as a function of $z$ for redshifts larger
      than 0.75 (see text for further details). Symbols with different
      colors refers to different galaxy clusters in each separate
      redshift bin. Dashed black-red line is the model given by
      \citet{song11}.  The solid green lines are the best fit to the
        simulated red-sequence relation used in this work and dashed
        green lines enclose $68\%$ of the galaxies. The
      area between them represents the color space used for the
      selection of galaxies described in Sect. \ref{sec:selection}.}
  \label{fi:CMR} 
\end{figure*}

We model the selection carried out in real world circumstances by
following the procedure we have developed for the South Pole Telescope
(SPT) dynamical mass calibration program (\citealt{bazin12}).  Namely,
we preferentially choose the most luminous red sequence galaxies that
lie projected near the center of the cluster for our spectroscopic
sample.  To do this we select galaxies according to their colors,
which are a direct prediction of the adopted semi-analytic model. In
particular, we compute the following color-magnitude diagrams for
different redshift ranges: $g-r$ as a function of $r$ for redshift
z$\leq 0.35$, $r-i$ as a function of $i$ for redshifts $0.35<$z$\leq
0.75$ and $i-z$ as a function of $z$ for redshifts larger than 0.75
(e.g. \citealt{song11}).  We report in Fig.~\ref{fi:CMR} the
color-magnitude diagram at different redshifts for all the galaxies
within the virial radius of each cluster. The model given by
\citet{song11}, which has proven to be a good fit to the
  observational data, is highlighted with a dashed black-red line. As
it is shown, the simulated cluster galaxy population has a
red-sequence flatter than the observational results. Because the purpose
of this work is not to study the evolution of the cluster galaxy
population, but rather to see the effect of the selection of galaxies
on the estimated dynamical mass, we adopt the following
procedure:   First we fit the red sequence at each analysed
redshift. Then, we symmetrically increase the area on
color-magnitude space in order to encompass $68\%$ of the galaxies and
iterate the fit.  The resulting best fit and corresponding area are
highlighted as green continuous and dashed lines in
Fig. \ref{fi:CMR}. Table \ref{t:cmr} describes the width in color space
used to select red sequence galaxies at each analysed redshift.
\begin{table} 
  \centering
  \caption{Color width of the simulated red sequence in
    magnitudes at each analysed redshift.
    Column 1: redshift; Column 2: 1 $\sigma$ width of the red sequence in magnitudes.}
  \begin{tabular}{llcr}
    z & mag \\  
    \hline 
0.00 &  0.05\\ 
0.09 &  0.06\\   
0.21 &  0.08\\
0.32 &  0.10\\
0.41 &  0.06\\
0.51 &  0.08\\
0.62 &  0.08\\
0.75 &  0.08\\
0.83 &  0.09\\
0.91 &  0.09\\
0.99 &  0.07\\
1.08 &  0.06\\
1.17 &  0.05\\
  \end{tabular}
  \label{t:cmr}
\end{table}

This color selection helps to reduce the interlopers in our cluster
spectroscopic sample.  In addition to color selection, we explore the
impact of imposing a maximum projected separation $R_\perp$ from the
cluster center, and we explore varying the spectroscopic sample size.
In all cases we use the $N_{gal}$ most massive (and therefore
luminous) galaxies in our selected sample. Table~\ref{tab:param} shows
the range of $N_{gal}$ and $a=R_\perp/r_{vir}$ that we explore as well
as the sample binning in redshift and mass.  Note that for SZE
selected clusters from SPT or equivalently X-ray selected samples of
clusters, once one has the cluster photometric redshift one also has
an estimate of the cluster virial mass and virial radius from SZE
signature or X-ray luminosity (e.g. \citealt{reiprich02},
\citealt{andersson11}); therefore, we do in fact restrict our
spectroscopic sample when building masks according to projected
distance from the cluster center relative to the cluster virial radius
estimate.

\begin{table} 
  \centering
  \caption{Parameter space explored for the mock observations. 
    Column 1: maximum projected distance $R_\perp$ from cluster 
    center $a=R_\perp/r_{vir}]$; Column 2: $N_{gal}$
    initial number of selected most massive red-sequence galaxies; Column 3:
    redshift $z$; Column 4: cluster mass \mvir [$10^{14} \msun$].}
  \begin{tabular}{llcr}
    $a$ & $N_{gal}$ & z & \mvir \\  
    \hline 
    0.2    & 10  & 0.00 & 1.0 \\ 
    0.4    & 15  & 0.09 & 2.0 \\ 
    0.6    & 20  & 0.21 & 4.0  \\
    0.8    & 25  & 0.32 & 6.0 \\
    1.0    & 30  & 0.41 & 8.0 \\
    1.2    & 40  & 0.51 & 10.0 \\
    1.4    & 50  & 0.62 & 20.0 \\
    1.6    & 60  & 0.75 &  \\
    1.8    & 75  & 0.83 &  \\
    2.0    & 100 & 0.91 &  \\
    2.2    &     & 0.99 &  \\
    2.4    &     & 1.08 &  \\
           &     & 1.17 &  \\
  \end{tabular}
  \label{tab:param}
\end{table}

\subsection{Dynamical friction and Velocity Bias}
\label{sec:dynamical}
\begin{figure}
  \hbox{\psfig{file=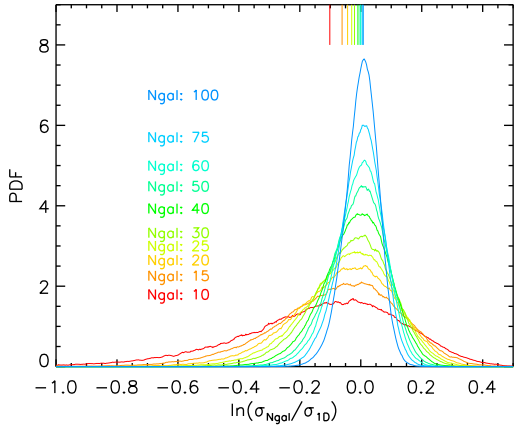,width=8.0cm}}
  \caption{ The probability distribution function of the measured velocity dispersion computed with different numbers of
    red-sequence cluster galaxies sorted by luminosity and normalized by the intrinsic 1D velocity dispersion using the full galaxy sample. The position of the mean of each curve is highlighted with a vertical line segment at the top of the figure.  Small samples of the most luminous galaxies exhibit biases in dispersion and significant asymmetries in the PDF.}
  \label{fi:dynfric}
\end{figure}

In section \ref{sec:intrinsic} we showed the presence of a tight
relation between the 3D dynamical mass and the virial mass \mvir\ for
galaxy clusters. When dynamical masses are computed from the 1D
velocity dispersion instead of the 3D one, we significantly increase
the scatter of this relation and introduce a negligible bias
($\lesssim 1\%$) due to the triaxial properties of dark matter
halos. We now study the effect of velocity segregation due to
dynamical friction and its effect on the estimated dynamical
masses. To do this, for each cluster we select a number of
red-sequence galaxies within the virial radius \rvir that ranges from 10
to 100 galaxies as described in Table~\ref{tab:param}.  We sort
galaxies according to their luminosity (different bands were used at
different redshift as described in Sec. \ref{sec:selection}). This
results in a "cumulative" selection. Therefore, for example, for each
cluster the 10 most luminous red-sequence galaxies are present in all
the other samples with larger number of galaxies.  On the other
hand, when a cluster field is spectroscopically observed,
completeness to a given limiting magnitude is not always achieved. 
In fact, the number of slits per mask in
multi-slit spectrographs is fixed, hence the central, high-density
regions of galaxy clusters can often be sampled only to brighter
magnitudes than the external regions. As a consequence, the spatial
distribution of the galaxies selected for spectroscopy turns out to
be more extended than the parent spatial distribution of cluster
galaxies. In the analyses presented here, we do not model this
observational limitation.  Indeed, as described
in the companion paper \citet{bazin12}, such difficulty could be easily
overcome by applying multiple masks to the same field, which would allow
one to achieve high completeness.  For each cluster and for all the three
orthogonal projections, we then compute the robust estimation of the
velocity dispersion \citet{beers90} with different numbers of galaxies
and compare it with the intrinsic 1D velocity dispersion. Fig.~\ref{fi:dynfric} 
shows the probability distribution
function (PDF) of the ratio between the velocity dispersion computed
with different numbers of bright red-sequence cluster galaxies
($\sigma_{Ngal}$) and the intrinsic 1D velocity dispersion
($\sigma_{1D}$) obtained by stacking results from all the lines of
sight of the cluster sample. Different colors refer to different
numbers of galaxies and the mean of each distribution is highlighted
at the top of the plot with a vertical line segment. We note that when
large numbers of galaxies are used to estimate the velocity
dispersion, the probability distribution function is well represented by a
log-normal distribution centered at zero. As a result dynamical masses obtained from
large numbers of bright red-sequence cluster galaxies are unbiased
with respect to the intrinsic 1D dynamical mass. However, when the
number of red-sequence galaxies used to estimate the velocity
dispersion is lower than $\sim 30$, the corresponding PDF starts to
deviate from a symmetric distribution and its mean is biased towards
lower values. This effect is evidence of a dynamically cold population
of luminous red galaxies whose velocities are significantly slowed due
to dynamical friction processes. Indeed dynamical friction is more
efficient for more massive galaxies, hence the velocity bias is
expected to be more important for the bright end of the galaxy
population (e.g. \citealt{biviano92}, \citealt{adami98},
  \citealt{cappi03}, \citealt{goto05}, \citealt{biviano06}).

\begin{figure*}
  \hbox{    \psfig{file=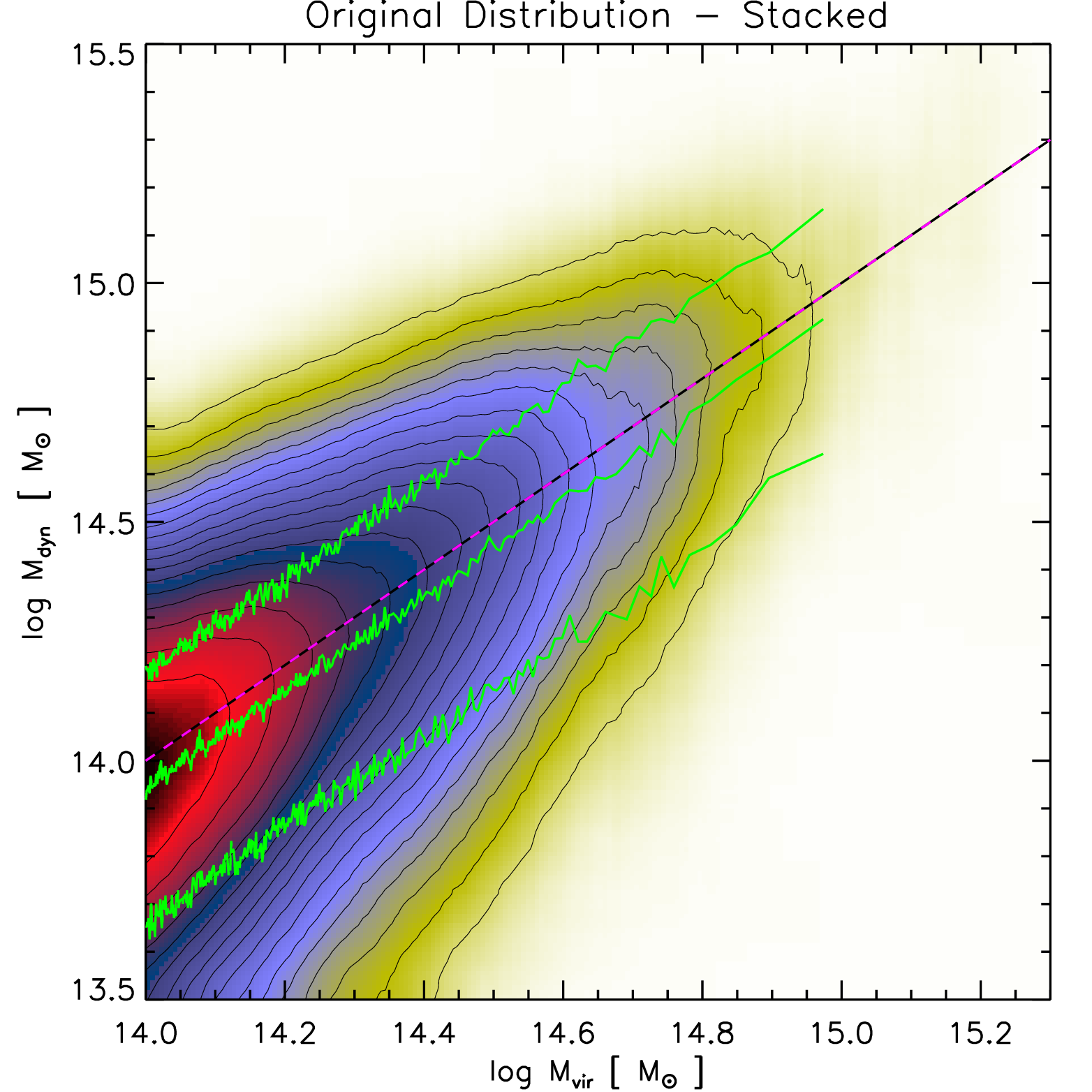,width=9.0cm}
    \psfig{file=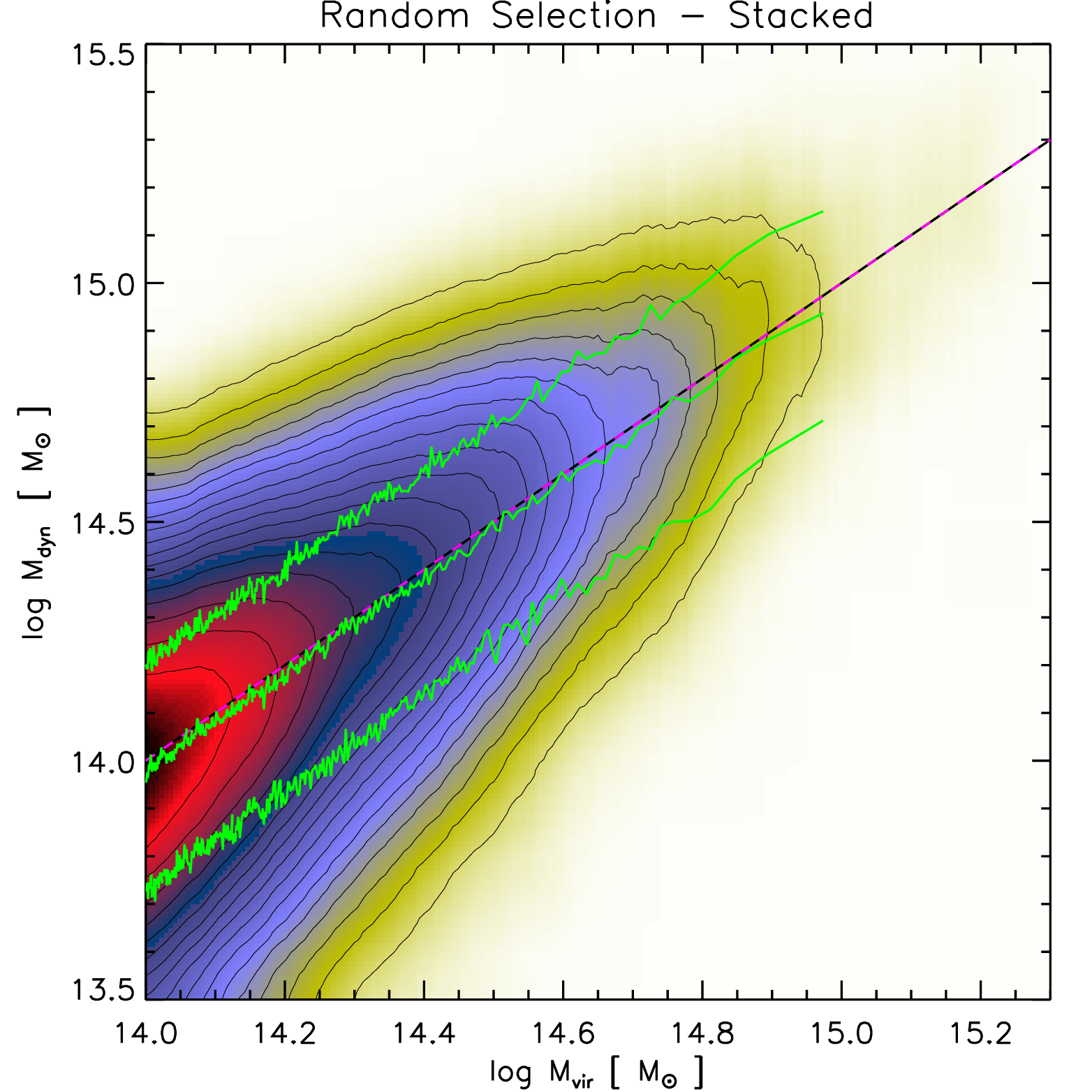,width=9.0cm}}
  \caption{ The relation between \mvir and the dynamical mass for all
    the clusters in the sample and for each orthogonal projection. For
    each cluster the dynamical mass is infer-ed by applying
    Eq. \ref{eq:fit} to the robust estimation of the velocity
    dispersion computed using different number of galaxies (Table
    \ref{tab:param}). Left panel is for bright red-sequence galaxies
    sorted according to their luminosity and right panel is for a
    randomly sorted array of galaxies. Dashed purple-black line is the
    one-to-one relation, and solid green lines represent the 16, 50 and
    84 percentiles.}
  \label{fi:fricran}
\end{figure*}

To verify this we compute $\sigma_{Ngal}$ in the same way described
above, but starting from galaxies that are randomly selected with
respect to luminosity. Note that in this case we only randomly select
galaxies, but we don't change the "cumulative nature" of our selection
and the subsequent estimated velocity dispersion when using larger
numbers of galaxies. We then calculate the corresponding dynamical
masses in the case of galaxies selected according to the procedure
described in Sect.~\ref{sec:selection} and in the case of random
selection using the different number of galaxies listed in
Table~\ref{tab:param}.  The resulting stacked dynamical masses for the
full sample of clusters and for the three orthogonal projections are
shown in Fig.~\ref{fi:fricran} as a function of the intrinsic virial
mass \mvir.  The dashed purple-black line is the one-to-one relation
and the solid green lines are the median and 16 and 84 percentiles of
the distributions. The left panel of Fig.~\ref{fi:fricran} represents
the original distribution, while the right panel represents the
randomly selected distribution. As expected from
Fig.~\ref{fi:dynfric}, if velocity dispersions are computed from
red-sequence galaxies selected according to their luminosity, a clear
bias is introduced in the estimated dynamical mass. Moreover we can
see that the distance between the median line and the 84 percentile
line is smaller than the distance between the median and the 16
percentile line, because the distribution is no longer symmetric.

\begin{figure*}
  \hbox{    \psfig{file=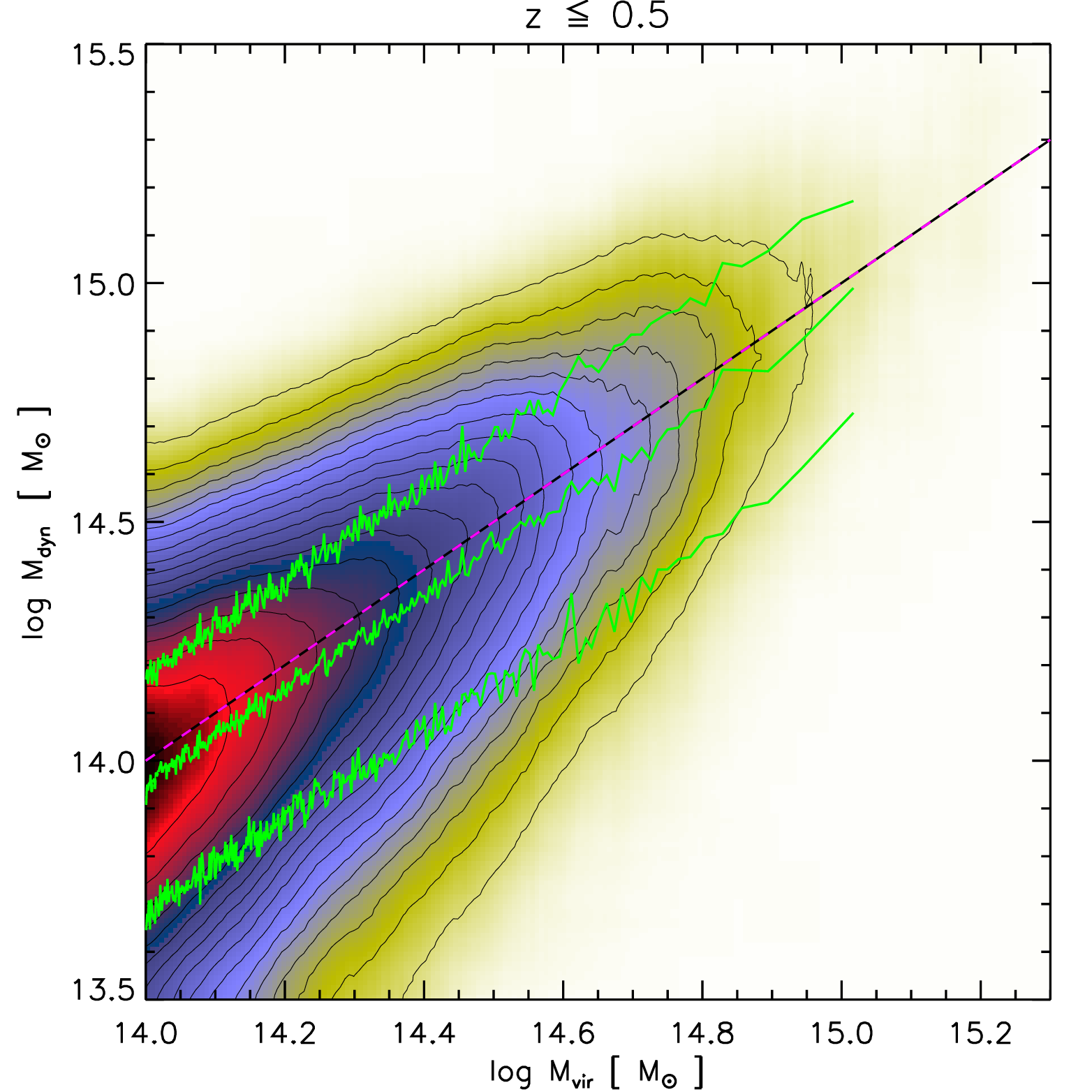,width=9.0cm}
    \psfig{file=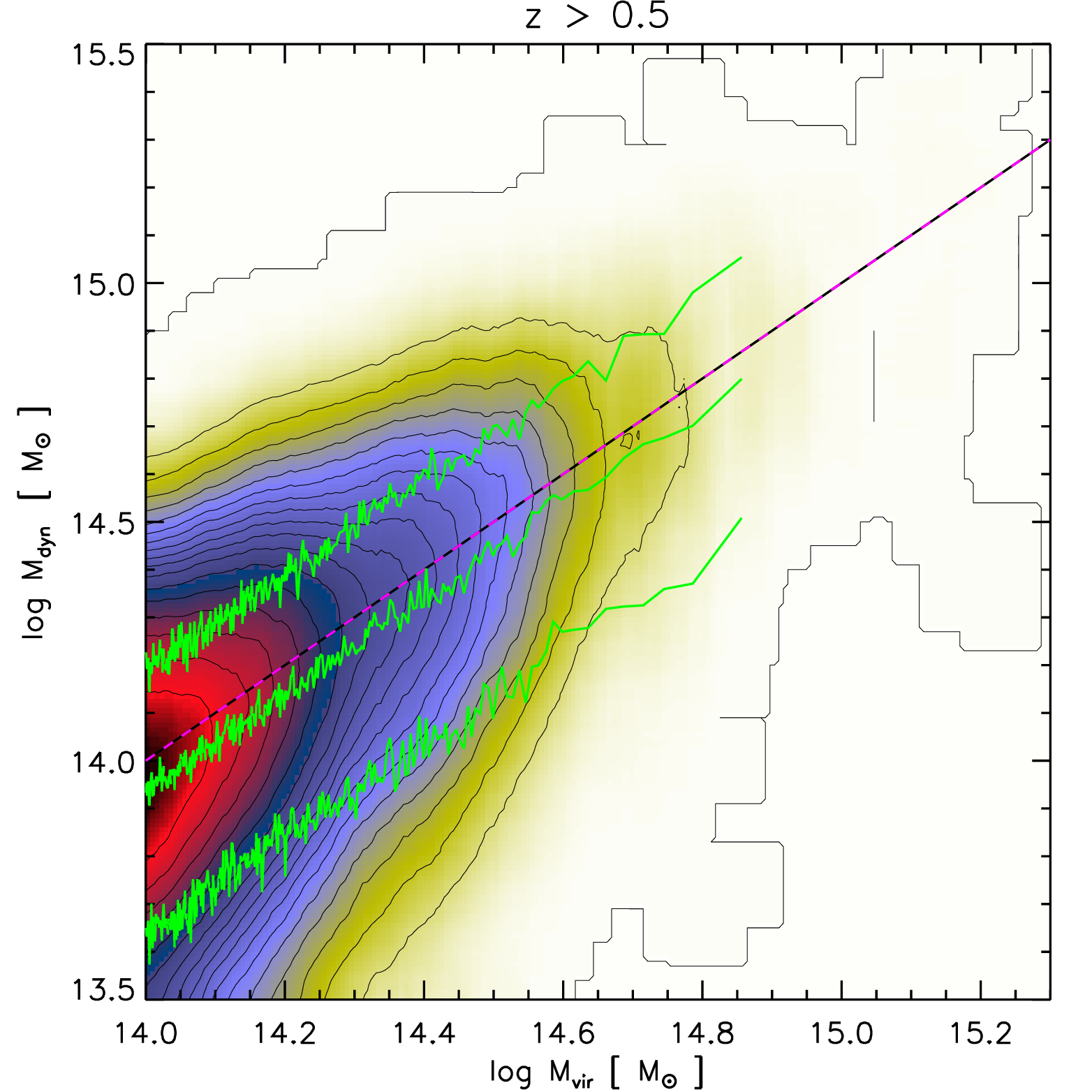,width=9.0cm}}
  \caption{ Same as for the left panel of Fig. \ref{fi:fricran}, but
    dividing our cluster sample in 2 redshift bins. Left panel is for
    $z\leq 0.5$ and right panel is for $z > 0.5$.}
  \label{fi:fricz}
\end{figure*}

Furthermore it appears that the bias
present in the estimated dynamical mass does not depend on the cluster
mass. On the other hand, if we randomly select galaxies (right
panel), the bias is reduced, and we obtain a more
symmetric distribution. We also check for a possible redshift
dependency on the velocity bias or dynamical friction. For this
purpose we split our sample of clusters into two different redshift
bins and show separately in Fig.~\ref{fi:fricz} the relation between
the true cluster virial mass and the estimated dynamical masses
computed with different number of bright red-sequence galaxies
selected according to their luminosity in the case of low redshift
(left panel) and high redshift (right panel) clusters. Obviously the
number of clusters and their mass distribution is a strong function of
redshift. However, it is worth noting that the impact of dynamical
friction on the estimation of velocity dispersion and dynamical mass
does not vary much with cluster mass or redshift.

Using the results of these mock observations we express both the
velocity dispersion bias, represented by the position of the vertical
segment at the top of Fig.~\ref{fi:dynfric}, and the characteristic
width of each distribution shown in Fig.~\ref{fi:dynfric} with the
following parametrisation: \be <ln(\sigma_{Ngal}/\sigma_{1D})> = 0.05
- 0.51/\sqrt {Ngal},
\label{eq:meandyn}
\ee
\be
\sigma_{ln(\sigma_{Ngal}/\sigma_{1D})} = -0.037 + 1.047/\sqrt {Ngal}.
\label{eq:sigmadyn}
\ee 
This parametrisation is valid only in the limit of the number of galaxies used in this study (between 10 and
100). For example if the dynamical mass is estimated starting from a
number of galaxies larger than 100, the bias would presumably be zero
rather than negative as implied from Eq.~\ref{eq:meandyn}.

\begin{figure*}
  \hbox{\psfig{file=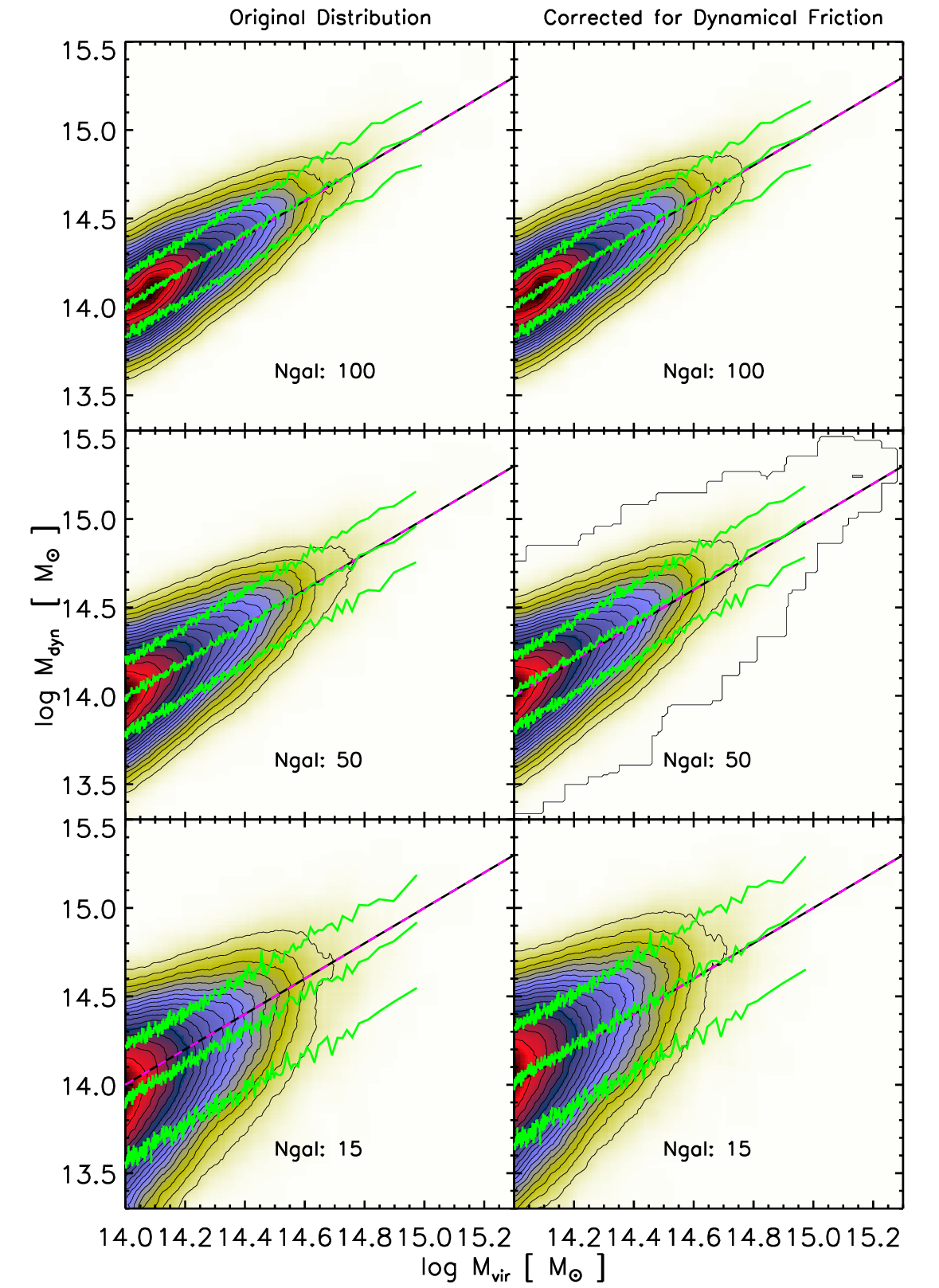,width=16.0cm}}
  \caption{ The relation between \mvir and the dynamical mass for all
    the clusters in the sample and for each orthogonal
    projection. {\it Left panels}: For each cluster the dynamical mass
    is infer-ed by applying Eq. \ref{eq:fit} to the robust estimation
    of the velocity dispersion computed using the 100, 50 and 15 most
    luminous red-sequence cluster galaxies (with distance from the
    centre smaller than \rvir) and show respectively in the upper,
    middle and lower panels. {\it Right panels:} Same as for left
    panels, but velocity dispersions are corrected according to
    Eq. \ref{eq:meandyn}. Dashed purple-black line is the one-to-one
    relation, and solid green lines represent the 84, 50 and 16
    percentiles.}
  \label{fi:dynfricpage}
\end{figure*}

We demonstrate in Fig.~\ref{fi:dynfricpage} that by applying
Eq.~\ref{eq:meandyn} to the velocity dispersion estimated with
different numbers of red-sequence cluster galaxies we are able to
remove the bias induced by the dynamical friction. In particular,
Fig.~\ref{fi:dynfricpage} shows the relation between true virial mass
and the dynamical mass estimated using the most luminous 100, 50 and
15 red-sequence galaxies.  Dynamical masses are computed by
applying Eq.~\ref{eq:fit} directly to the velocity dispersion (left
panels) and to the velocity dispersion corrected according to
Eq.~\ref{eq:meandyn} (right panels). Dynamical friction is affecting
mostly the bright end of the red-sequence cluster galaxies population
and therefore the bias is larger in the case of the smallest sample (lower left
panel). Consistently the correction given by Eq.~\ref{eq:meandyn} is
larger in this case, whereas it is negligible in the other cases (50 and 100).

\subsection{Impact of Poisson Noise}
\label{sec:poisson}

\begin{figure}

  \hbox{    \psfig{file=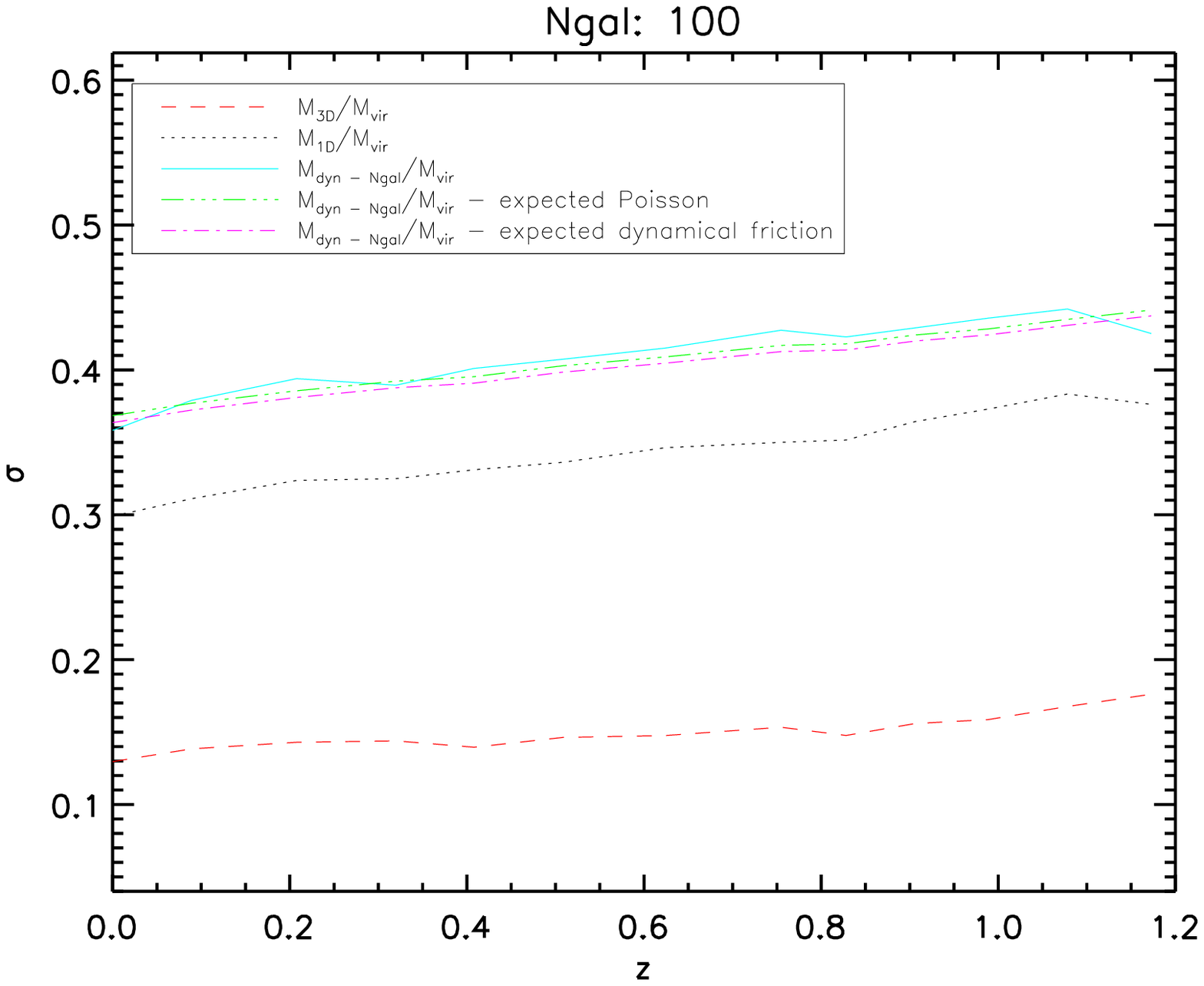,width=9.0cm}}
  \caption{ The evolution of the 1 $\sigma$ scatter as a function of
    redshift in log-space for the following quantities. Dashed red
    (dotted black) line is for the ratio between the estimated
    dynamical mass $M_{3D}$ ($M_{1D}$) computed from the 3D (1D)
    velocity dispersion and the virial mass \mvir.  The solid cyan
    line is for the ratio between the measured dynamical mass
    $M_{dyn}$ computed from the 100 most luminous red-sequence
    galaxies within the virial radius of each cluster along each line
    of sight and the virial mass \mvir.  The dashed-dotted green line
    is the expected scatter in mass obtained by multiplying $B$ by the
    term given by adding in quadrature the scatter from the 1D
    velocity dispersion and a Poissonian term equal to $\sqrt{2\times
      100}$. The dashed-dotted purple line is the expected scatter in
    mass obtained by multiplying $B$ by the term given by adding in
    quadrature the scatter from the 1D velocity dispersion and a term
    computed with the fitting formula of Eq.~\ref{eq:sigmadyn}.}
  \label{fi:dynfric100}
\end{figure}

In this work we restrict our analyses to all the galaxies with stellar
masses predicted by the adopted SAM larger than $5\times 10^8
\msun$. The total number of galaxies within the virial radius
\rvir\ is therefore quite large and even for the poorer clusters with
$\mvir \sim 10^{14} \msun$, the number of galaxies used to compute the
1D velocity dispersion is of the order of $N_{1D} \sim 200$.  As a
result, in the absence of any dynamical friction effect, the
associated characteristic scatter to the ratio $\sigma_{Ngal}/
\sigma_{1D}$ is well represented by the Poissonian factor
$\sqrt{2N_{gal}}$. To demonstrate it, we show in
Fig.~\ref{fi:dynfric100} the evolution of the scatter in the relation
between the true virial masses and the dynamical masses as a function
of redshift. For each cluster, dynamical mass is
estimated starting from the velocity dispersion of the 100 most
luminous red-sequence galaxies through Eq.~\ref{eq:fit}. The resulting
scatter is highlighted as a cyan solid line. We also show the
evolution of associated scatter when dynamical mass is computed from
the intrinsic 3D (dashed red line) and 1D (dotted black line) velocity
dispersions. Moreover, similarly to Fig.~\ref{fi:stddev}, we
separately show the predicted scatter obtained by adding in quadrature
the scatter associated to the 1D velocity dispersion with the Poisson
term $\sqrt{2 N_{gal}}$ (dashed-triple dotted green line) or with the
factor given by Eq.~\ref{eq:sigmadyn} (dashed-dotted purple line). We
note that, as expected, both predictions agree very well with the
measured evolution of the scatter.

\begin{figure*}
  \hbox{    \psfig{file=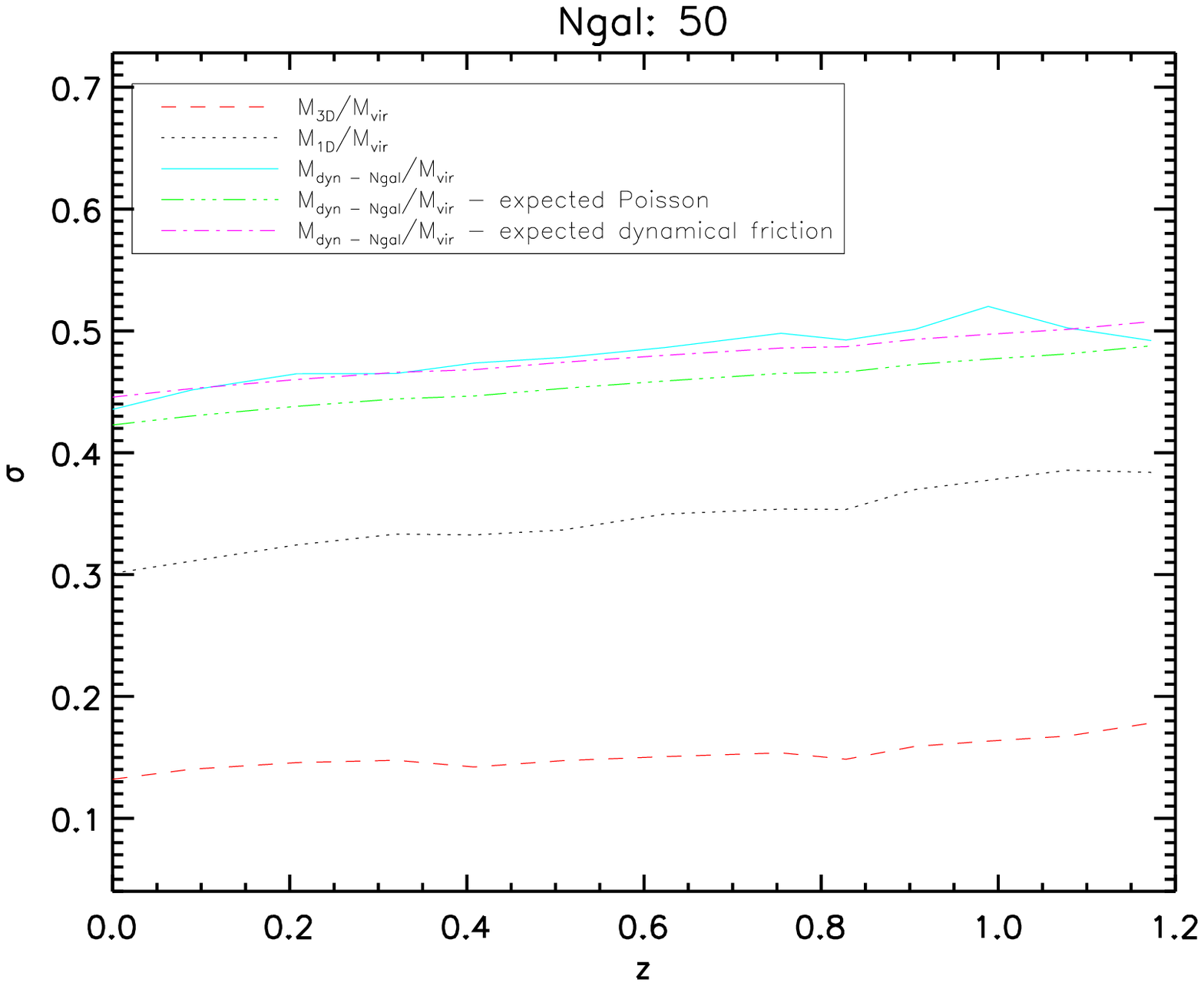,width=9.0cm}
    \psfig{file=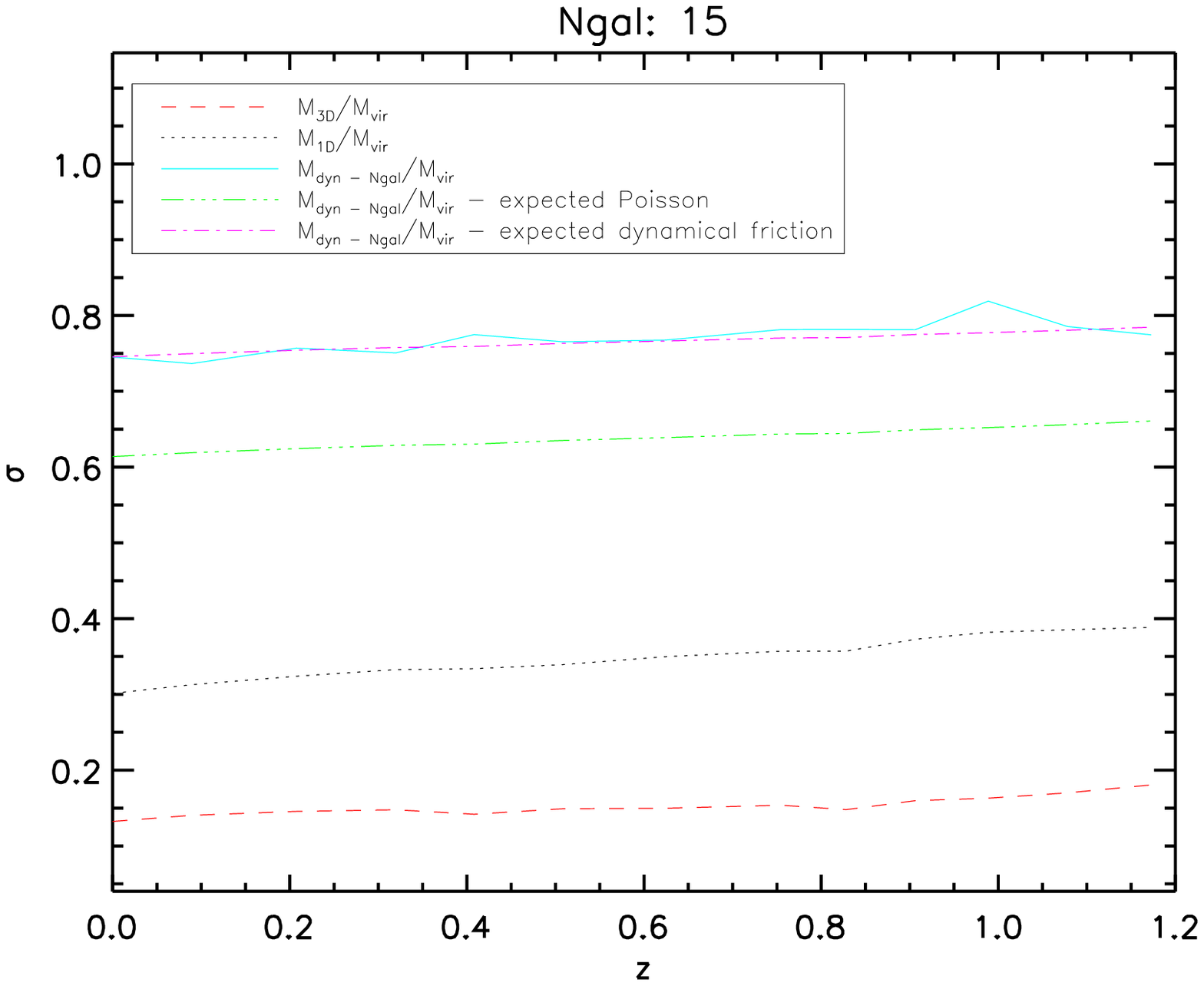,width=9.0cm}}
  \caption{ Same as for Fig. \ref{fi:dynfric100}, but with a number of galaxies respectively equal to 50 (left panel) and 15 (right panel).}
  \label{fi:dynfric5015}
\end{figure*}

However, if a lower number of galaxies is used to calculate the
dynamical mass, a difference in the two predictions emerge. For
example, in Fig.~\ref{fi:dynfric5015} we show the same computation
highlighted in Fig.~\ref{fi:dynfric100}, but with a number of galaxies
equal to 50 (left panel) and 15 (right panel). We note in particular
that the observed evolution of the scatter of the relation among the
virial mass and the dynamical mass is well described by adding in
quadrature to scatter associated to the intrinsic 1D dynamical mass
the term given by the fitting formula of Eq.~\ref{eq:sigmadyn}. On the
contrary, if only the Poisson term $\sqrt{2 N_{gal}}$ is taken into
account, the predicted scatter is underestimated with respect to the
measured one. Furthermore, note that while on the one hand in
Figures~\ref{fi:dynfricpage} and \ref{fi:dynfric} we showed that the dispersion
calculated using 50 galaxies does not introduce a significant mass bias;
however, it is clear that the Poisson term is no
longer adequate to explain the real scatter.  On the other hand,
if the contribution to the scatter due to dynamical friction is
included through Eq. \ref{eq:sigmadyn} we are able to recover the
right answer.

%%%%%%%%%%%%%%%%%%%%%%%%%%%%%%%%%%%%%%%%%%%%%%%%%%%%%%%%%%%%%%%%%%%%%%%%%%%%%%%
\subsection{Impact of Interlopers}
\label{sec:Interlopers}

\begin{figure}
  \hbox{\psfig{file=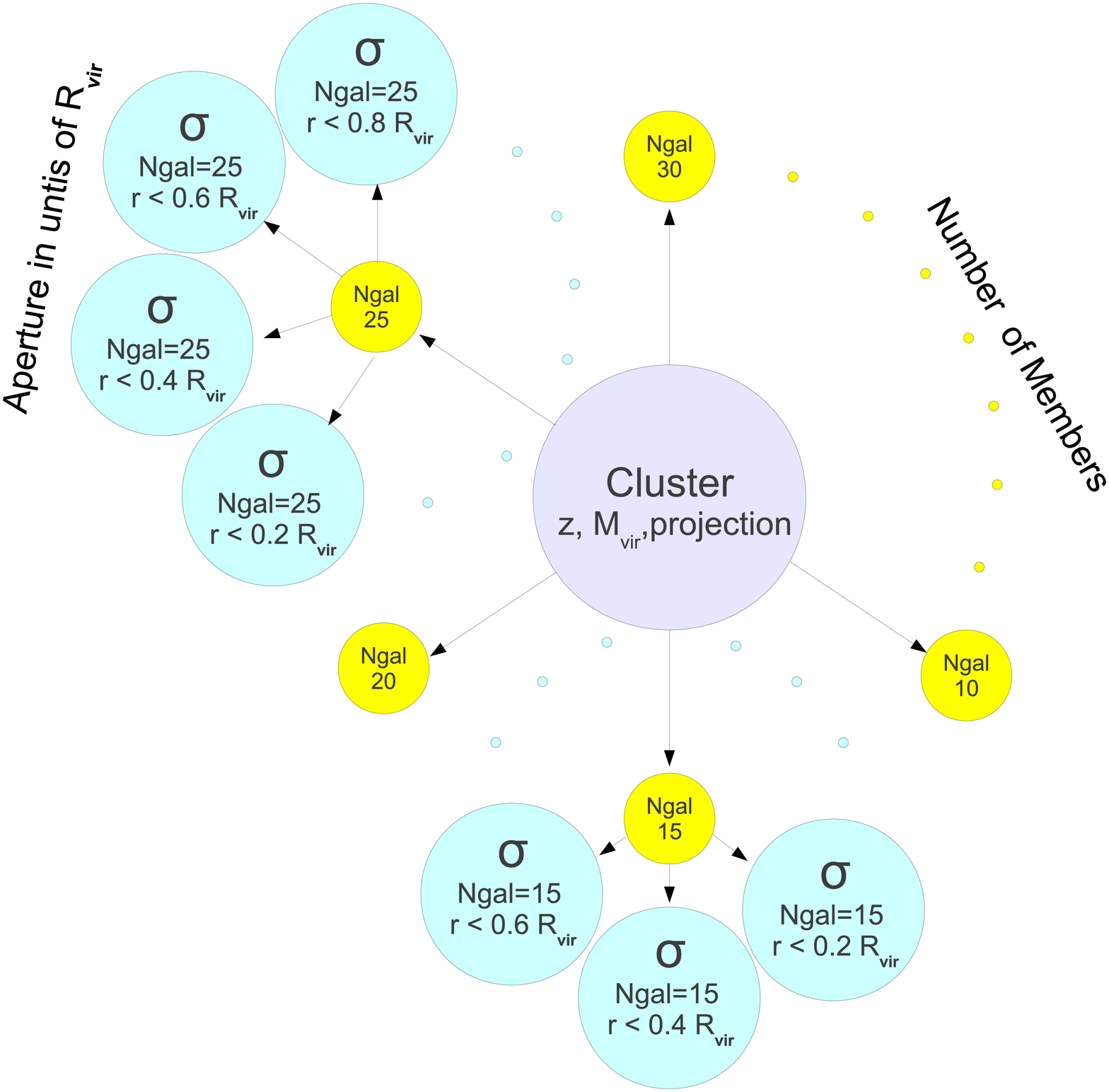,width=9.0cm}
    }
  \caption{ Schematic representation of the parameter space explored
    in this work.  See Table~\ref{tab:param} for specific ranges in each parameter.}
  \label{fi:schema}
\end{figure}

Finally, to have a more coherent and realistic approach to our
analyses, we further investigate the effect of interlopers as a
possible source of systematics in the computation of clusters
dynamical mass. For this purpose, for each snapshot and projection, we
construct a number of cylindrical light-cones centred at each cluster
with height equal to the full simulated box-length and different
radius spanning the interval 0.2 to 2.4 \rvir. The different aperture
values used are listed in Table~\ref{tab:param}. We then apply an
initial cut of $4000 \vel$ to select galaxies within the
cylinders. For each cylindrical light-cone realisation, we then
initially select a different number of members in the color-magnitude
space described in Sect. \ref{sec:selection} ranging from 10 to 100
galaxies as shown in Table \ref{tab:param}.  Several techniques have
been developed to identify and reject interlopers. Such methods have
been studied before typically using randomly selected dark matter
particles \citep[e.g.][]{perea90,diaferio99,lokas06,wojtak07,wojtak09}
and more recently using subhalos by \citet{biviano06} and
\citet{white10}. However, for the purpose of this work, here we simply
apply a 3$\sigma$ clipping procedure to its robust estimation of the
velocity dispersion \citet{beers90} to reject interlopers, as
discussed in \citet{bazin12}. This leads to a final spectroscopic
sample of galaxies for each cluster, at each redshift, for each
projection, within each different aperture and for each different
initially selected number of red-sequence galaxies.
Fig.~\ref{fi:schema} is a schematic representation of the procedure we
follow to obtain from each cluster and projection different estimation
of the velocity dispersion according to different "observational
choices".

\begin{figure*}
  \hbox{\psfig{file=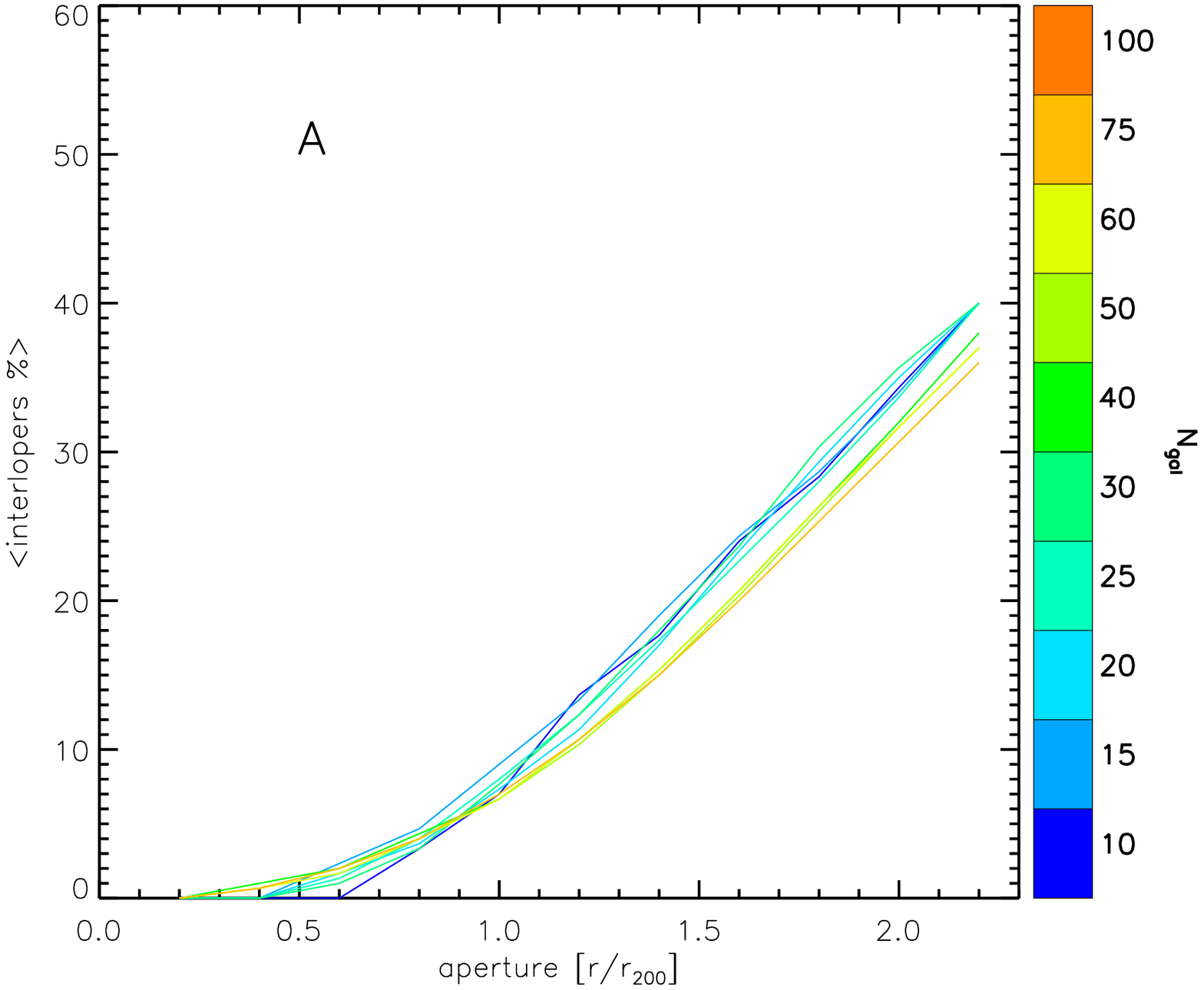,width=9.0cm}
    \psfig{file=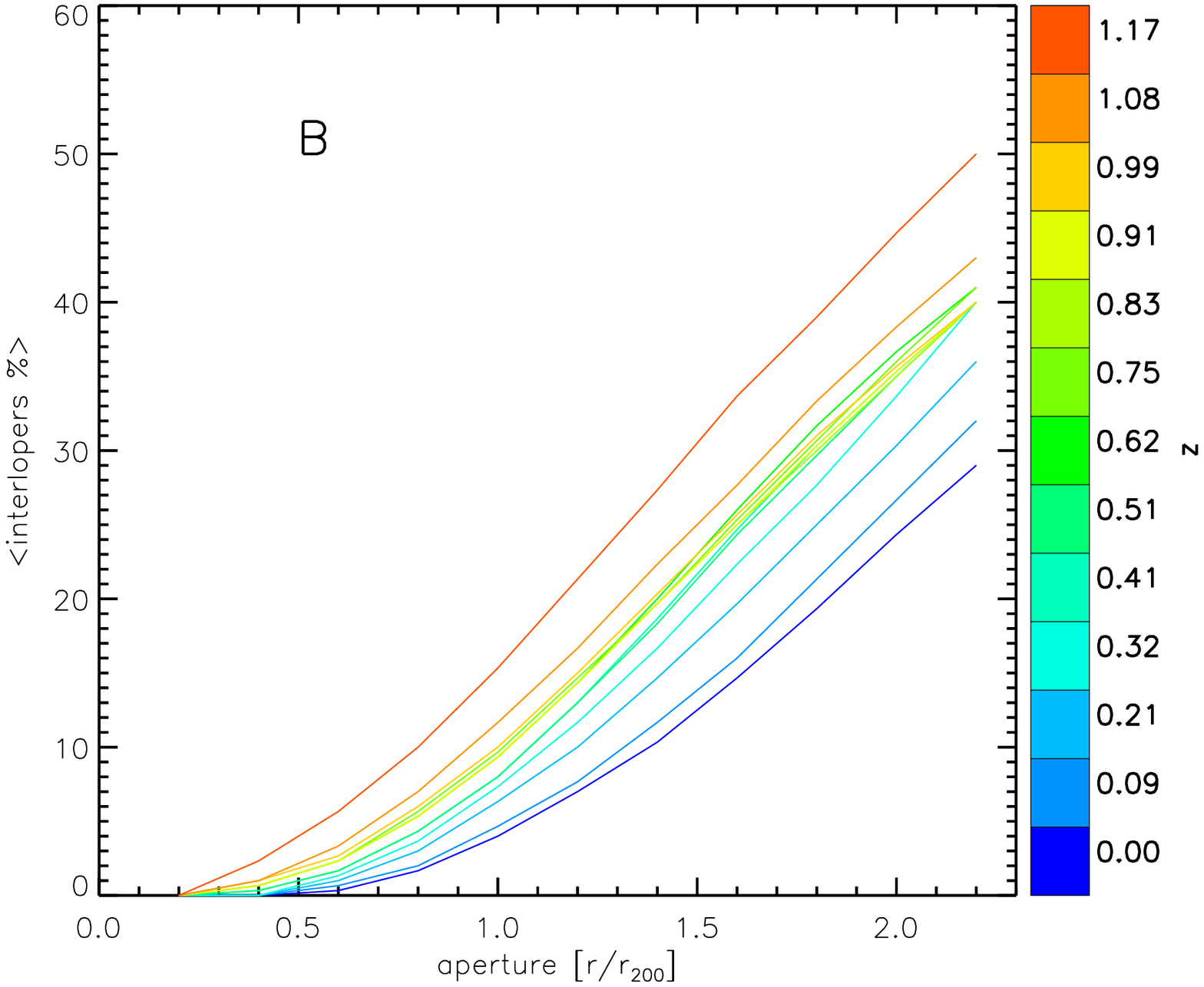,width=9.0cm}}
  \hbox{ \psfig{file=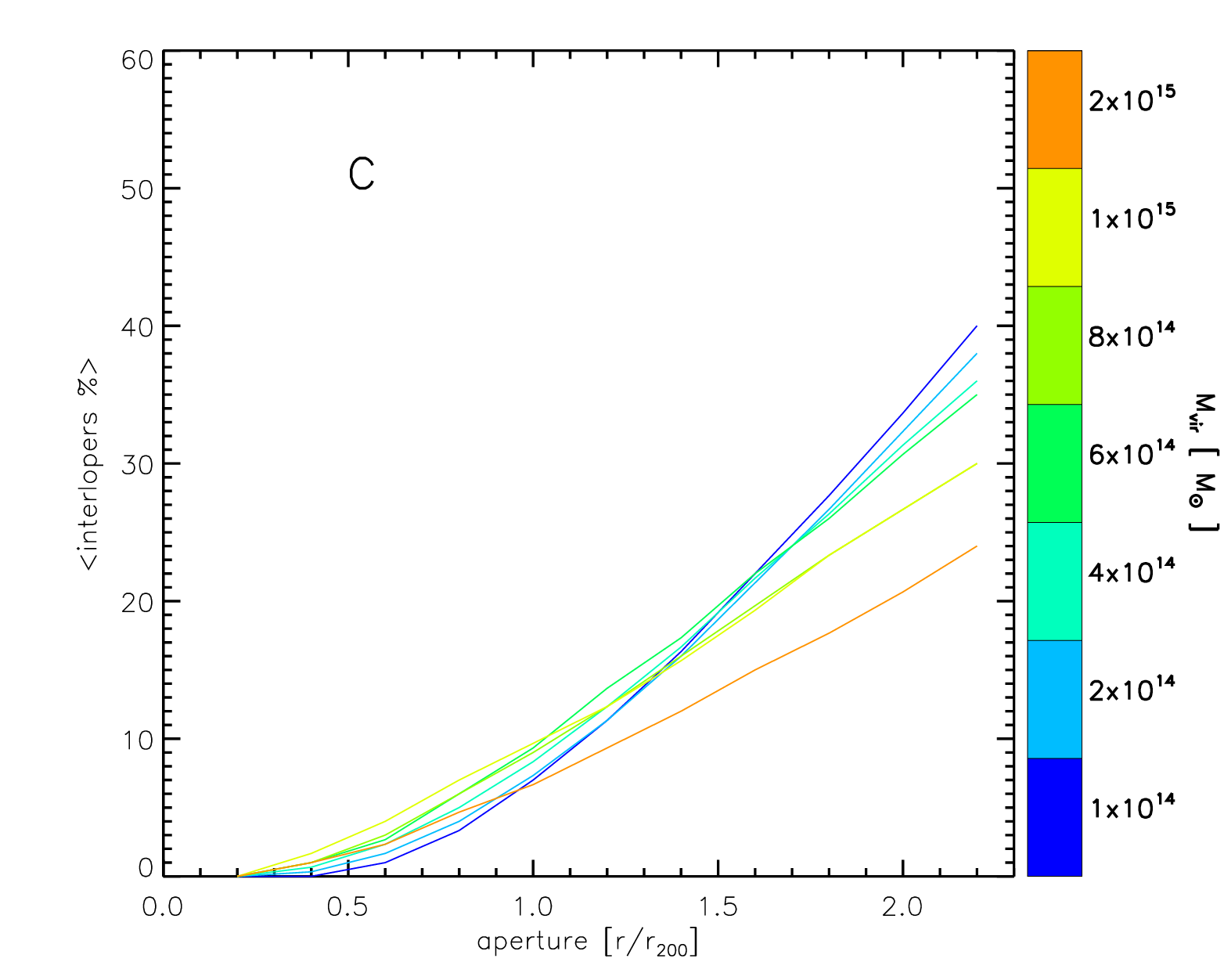,width=9.0cm}
    \psfig{file=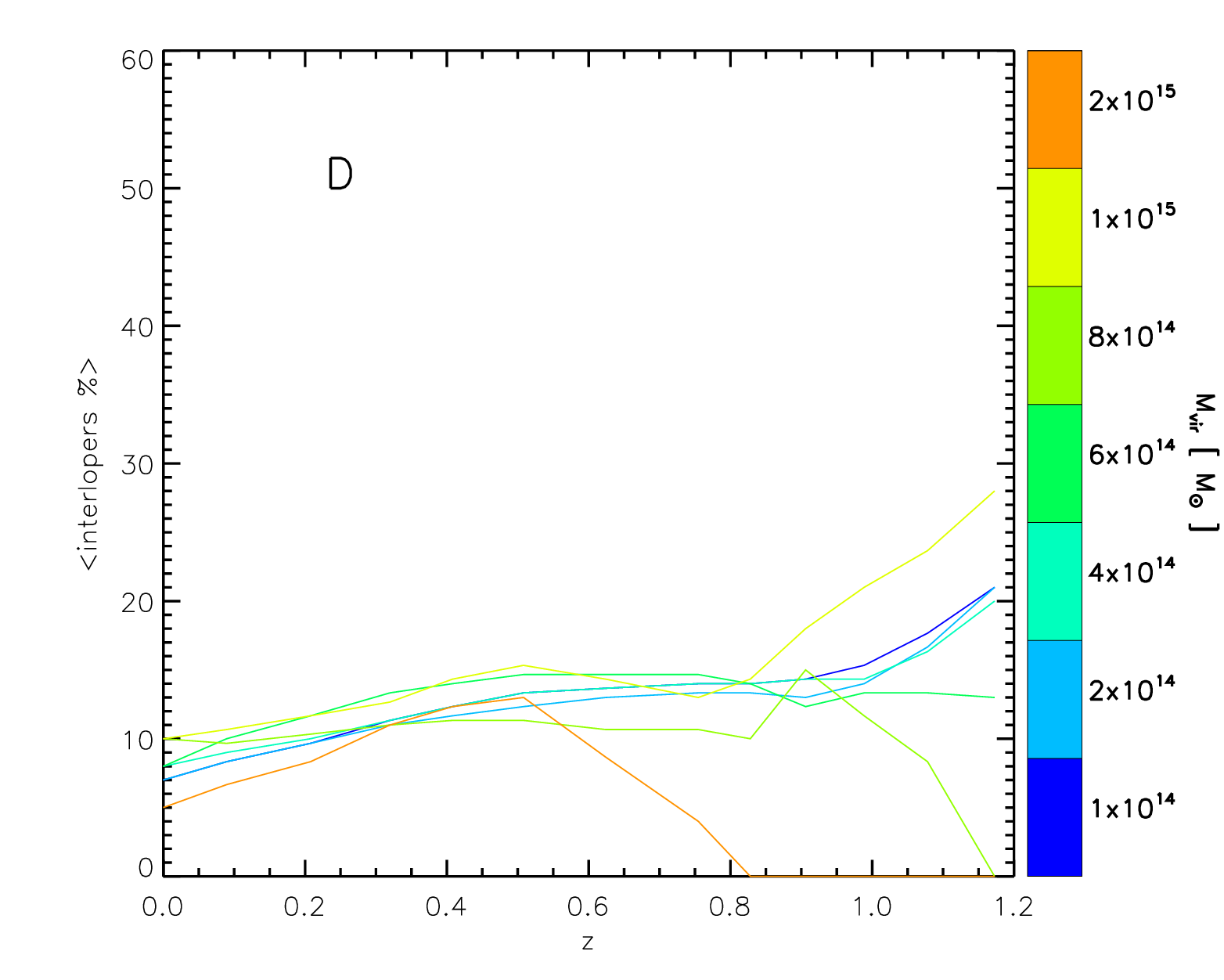,width=9.0cm}}
  \caption{ {\it Upper panels and bottom left panel:} The stacked mean
    fraction of interlopers (defined as galaxies at distance larger
    than $3\times$ \rvir) as a function of maximum projected
    separation from the cluster $R_\perp$ normalized to \rvir\ color
    coded in numbers of galaxies used to estimate the velocity
    dispersion (top left panel labeled A), redshift (top right panel
    labeled B) and mass of the cluster (bottom left panel labeled
    C). {\it Bottom right panel:} The stacked mean fraction of
    interlopers as a function of redshift color coded according to
    the cluster mass (labeled D).}
  \label{fi:interlopers}
\end{figure*}

From final spectroscopic sample of galaxies described above, we
compute the fraction of interlopers (arbitrary defined here as
galaxies lying at a cluster centric distance larger than
3$\times$\rvir) as a function of the aperture by stacking together the
sample in different bins according to their redshift, to the number of
galaxies used to evaluate their velocity dispersion and to the cluster
masses. This can be seen in Figure~\ref{fi:interlopers} and is in good agreement with previous works (e.g. \citealt{mamon10}). The two upper
panels and the lower-left panel show the fraction of interlopers as a
function of aperture respectively color coded according to the number
of galaxies (panel A), to the redshift (panel B) and to the cluster
mass (panel C). As expected, the fraction of interlopers rises with
the aperture within which the simulated red-sequence galaxies were
initially chosen. This indicates that even red sequence,
spectroscopically selected samples are significantly contaminated by
galaxies lying at distances more than three times the virial radius
from the cluster.

 On the other hand a much weaker dependency between the number of
 selected red sequence galaxies and the fraction of interlopers is
 highlighted on the upper-left panel of Fig. \ref{fi:interlopers}
 (A). Whether one has small or large samples the fraction of
 interlopers remains almost the same. The upper-right and the
 bottom-left panels are showing that the fraction of interlopers is
 larger at larger redshifts consistently with a denser Universe (B),
 and is a steeper function of aperture for lower mass clusters
 (C). Since in the hierarchical scenario more massive halos forms at
 later times than the lower mass ones, these two variables are clearly
 correlated. Thus, we also show in the bottom-right panel labeled "D"
 how the fraction of interlopers varies as a function of redshift by
 stacking together the sample in different mass bins. Most massive
 clusters are not formed yet at high redshifts, therefore above
 certain redshifts the redder lines go to zero. Although oscillating,
 an evident tendency of increasing fraction of interlopers is
 associated to larger redshift, whereas at fixed $z$ there is no clear
 dependency of the fraction of interlopers from the clusters mass. We
 stress however that all the relations shown in
 Fig.~\ref{fi:interlopers} are meant to describe the qualitative
 dependency of the interloper fraction from the analysed quantities.

\begin{figure*}
  \hbox{\psfig{file=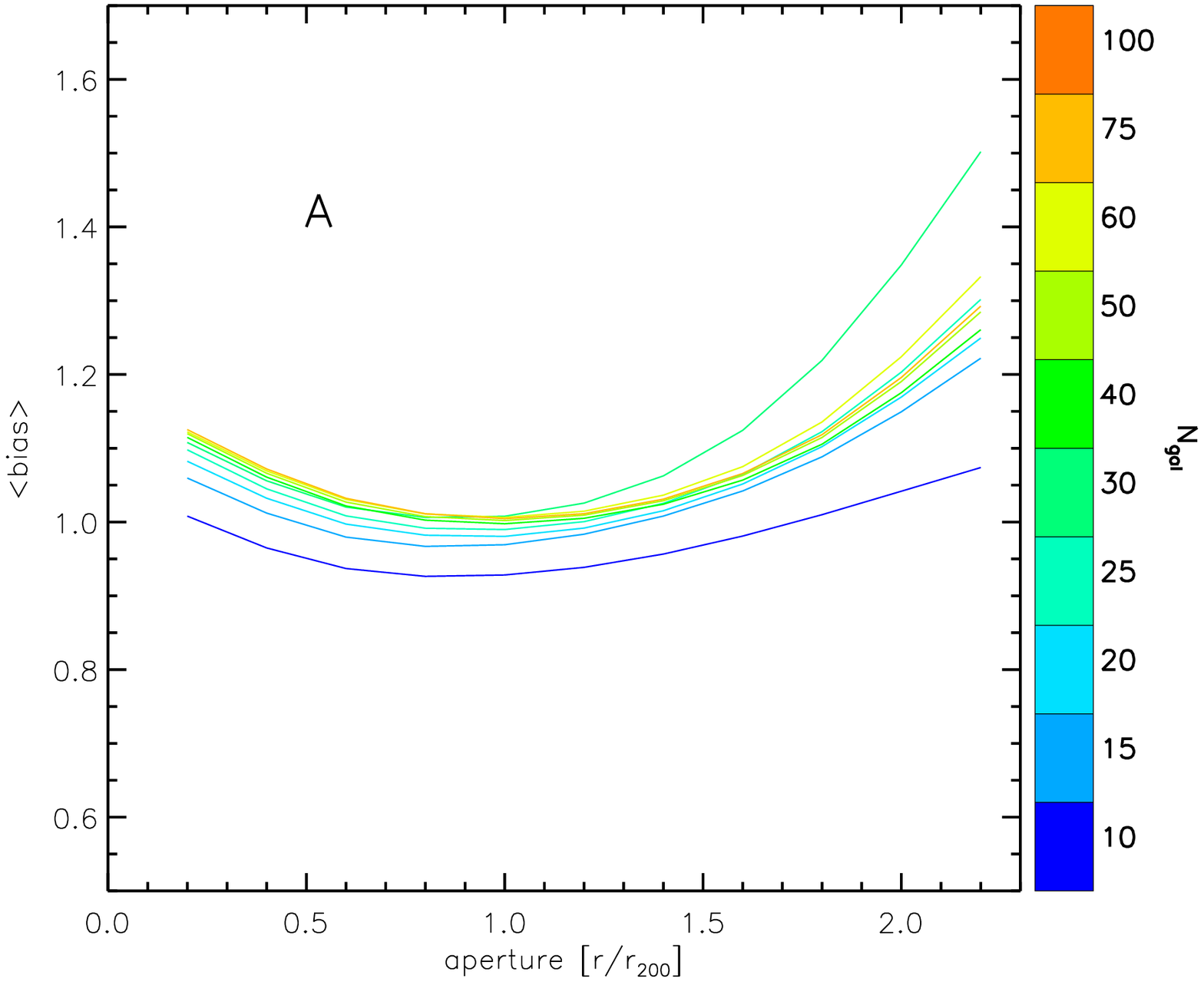,width=9.0cm}
    \psfig{file=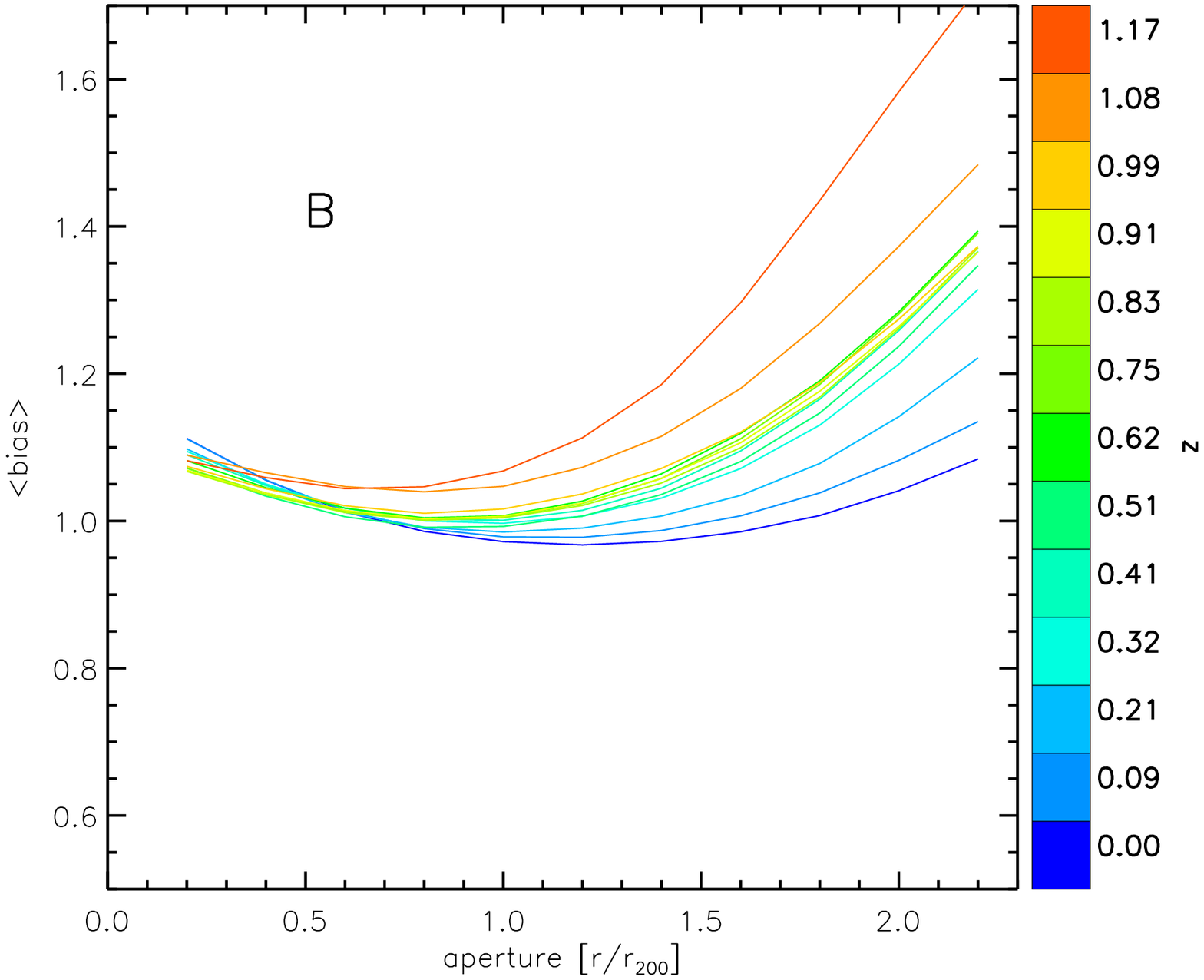,width=9.0cm}}
  \hbox{ \psfig{file=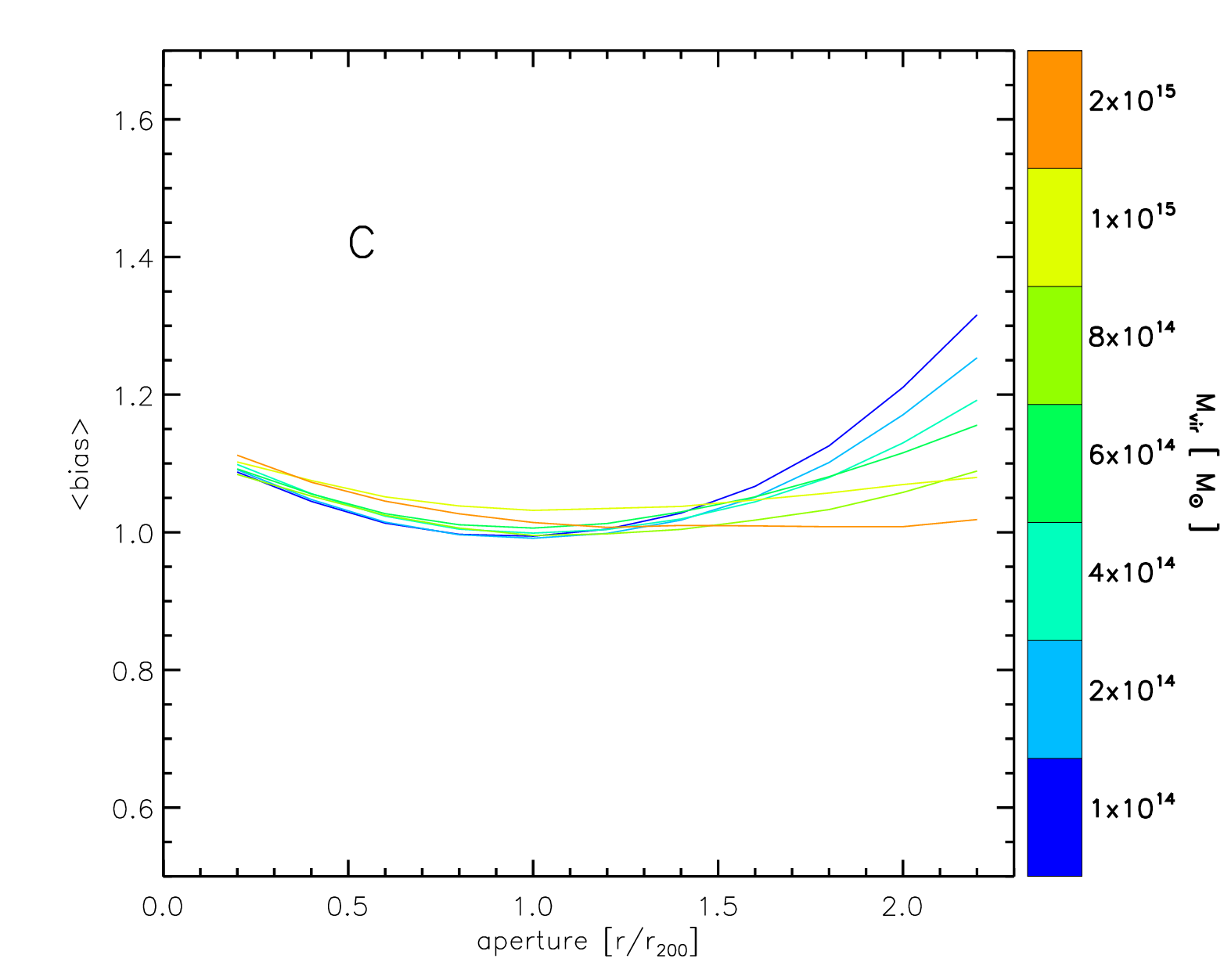,width=9.0cm}
    \psfig{file=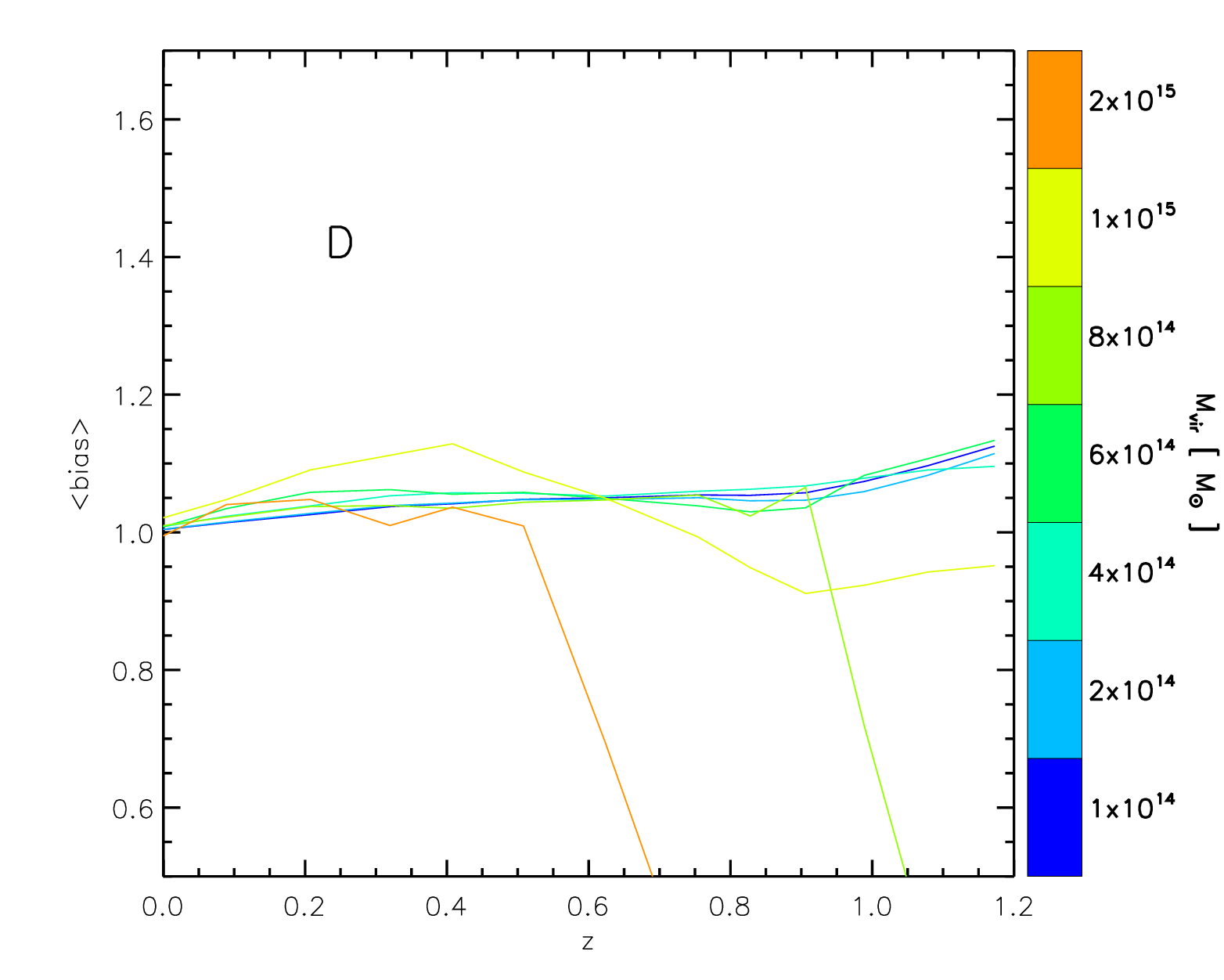,width=9.0cm}}
  \caption{ {\it Upper panels and bottom left panel:} The stacked mean
    velocity bias as a function of maximum projected separation from
    the cluster $R_\perp$ normalized to \rvir\ color coded in numbers
    of galaxies used to estimate the velocity dispersion (top left
    panel labeled A), redshift (top right panel labeled B) and mass of
    the cluster (bottom left panel labeled C). {\it Bottom right
      panel:} The stacked mean velocity bias as a function of redshift
    color coded according to the cluster mass (labeled D).}
  \label{fi:biasfrac}
\end{figure*}

\begin{figure}
  \hbox{\psfig{file=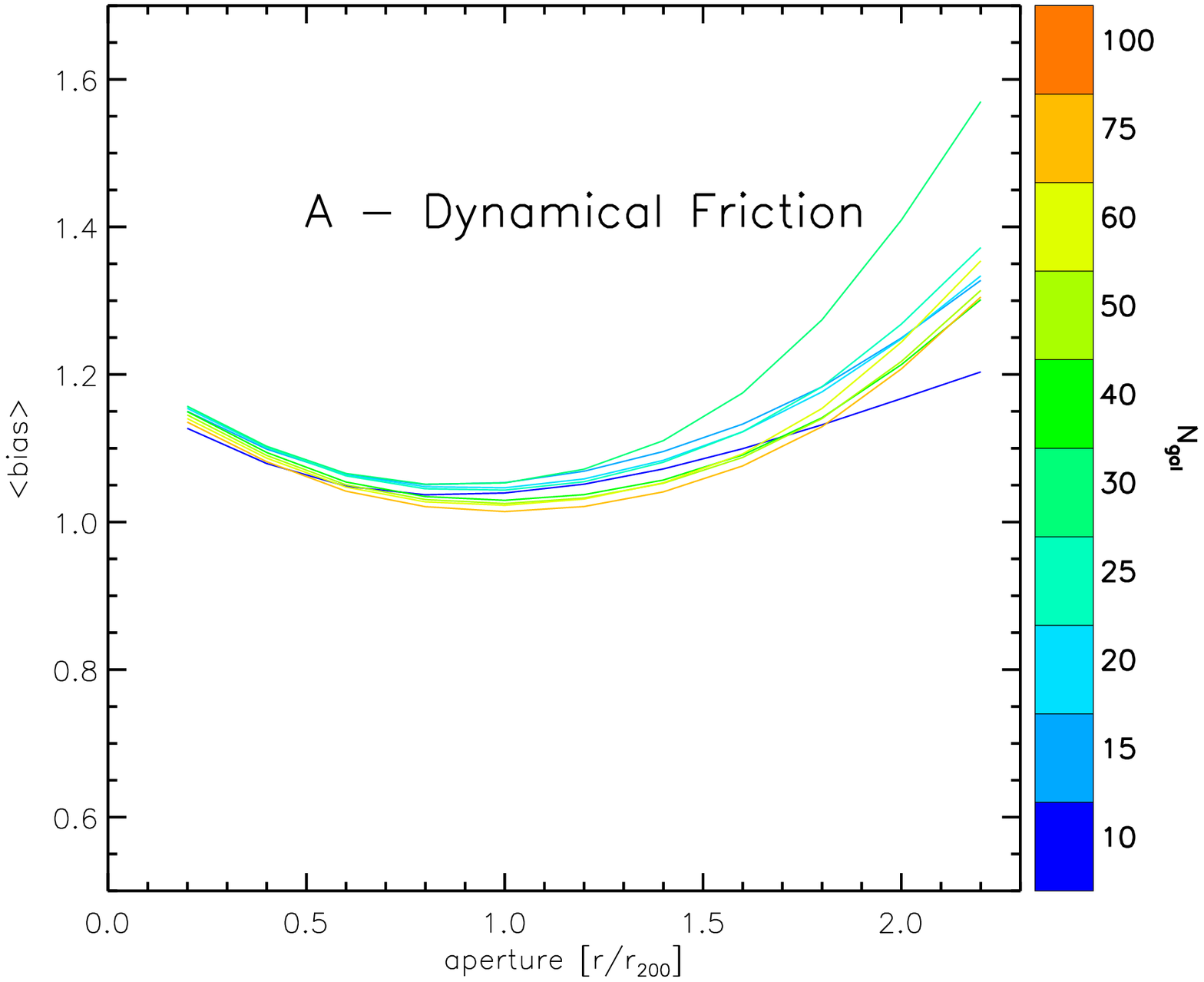,width=9.0cm}}
  \caption{ {\it Upper panels and bottom left panel:} Same as
    Fig. \ref{fi:biasfrac} - panel A, but corrected for dynamical
    friction according to Eq. \ref{eq:meandyn}.}
  \label{fi:biasdyn}
\end{figure}

In a similar way we compute the mean velocity bias (defined as
the ratio between the measured velocity dispersion and the intrinsic
line-of-sight velocity dispersion:
$\sigma_{(N_{gal},R_\perp,M_{vir},z)} / \sigma_{1D} $) as a function
of the aperture by stacking together the sample in different bins
according to their redshift, to the number of spectroscopic galaxies
and to the cluster mass. This can be seen in
Figure~\ref{fi:biasfrac}. The two upper panels and the lower-left
panel show the velocity bias as a function of aperture respectively
color coded according to the number of galaxies (panel A), to the
redshift (panel B) and to the cluster mass (panel C). Interestingly,
the velocity bias has a minimum when velocity dispersions are
evaluated within $R_\perp \sim \rvir$, and rises at both smaller and
larger radii.  In particular, for projected radii $\lesssim \rvir$,
where the effect of interlopers is smaller, we recover the expected
decrease of the average velocity dispersion profile
(e.g. \citealt{biviano06}) as a function of aperture. On the other
hand, for $R_\perp \gtrsim \rvir$, the larger contamination from
interlopers is significantly affecting and boosting the velocity
bias. Furthermore, as expected, dynamical friction is also affecting
the estimated velocity dispersion, when the latest is computed with
a small number of selected red sequence galaxies (A). Indeed, by
applying Eq.~\ref{eq:meandyn} to the estimated velocity dispersion
we are able to successfully remove the degeneracy between the
velocity bias and the number of galaxies within projected aperture
$R_\perp \lesssim \rvir$ (Fig. \ref{fi:biasdyn}). The upper-right and the
bottom-left panels of Fig.~\ref{fi:biasfrac} show that,
consistent with the fraction of interlopers, the velocity bias
computed within $R_\perp \gtrsim \rvir$ is larger at larger
redshifts (B) and is a steeper function of aperture for lower mass
clusters (C). Finally, a mild dependence of redshift for fixed mass
is highlighted in the bottom-right panel (D).

\begin{figure*}
  \hbox{\psfig{file=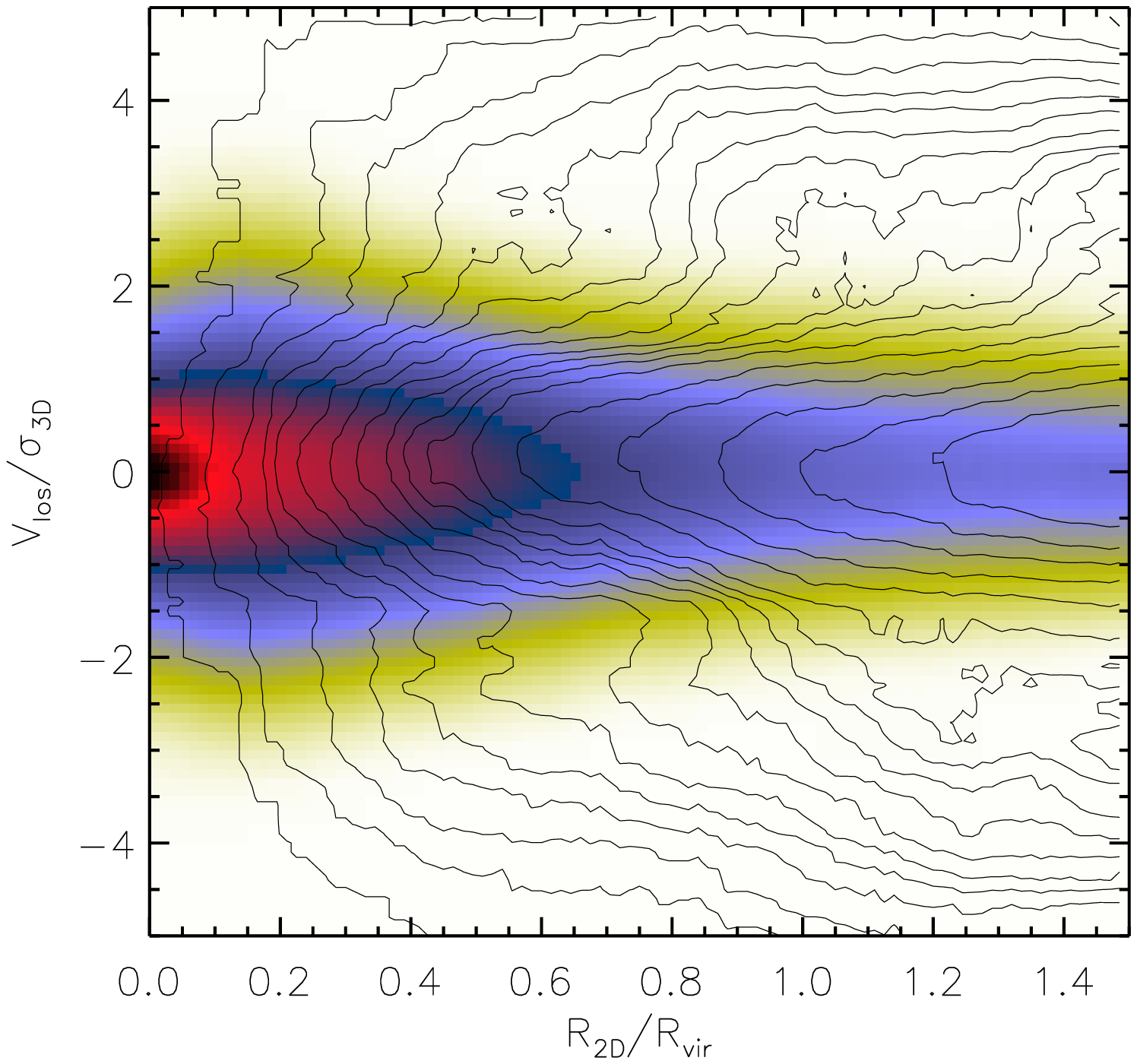,width=9.0cm}
    \psfig{file=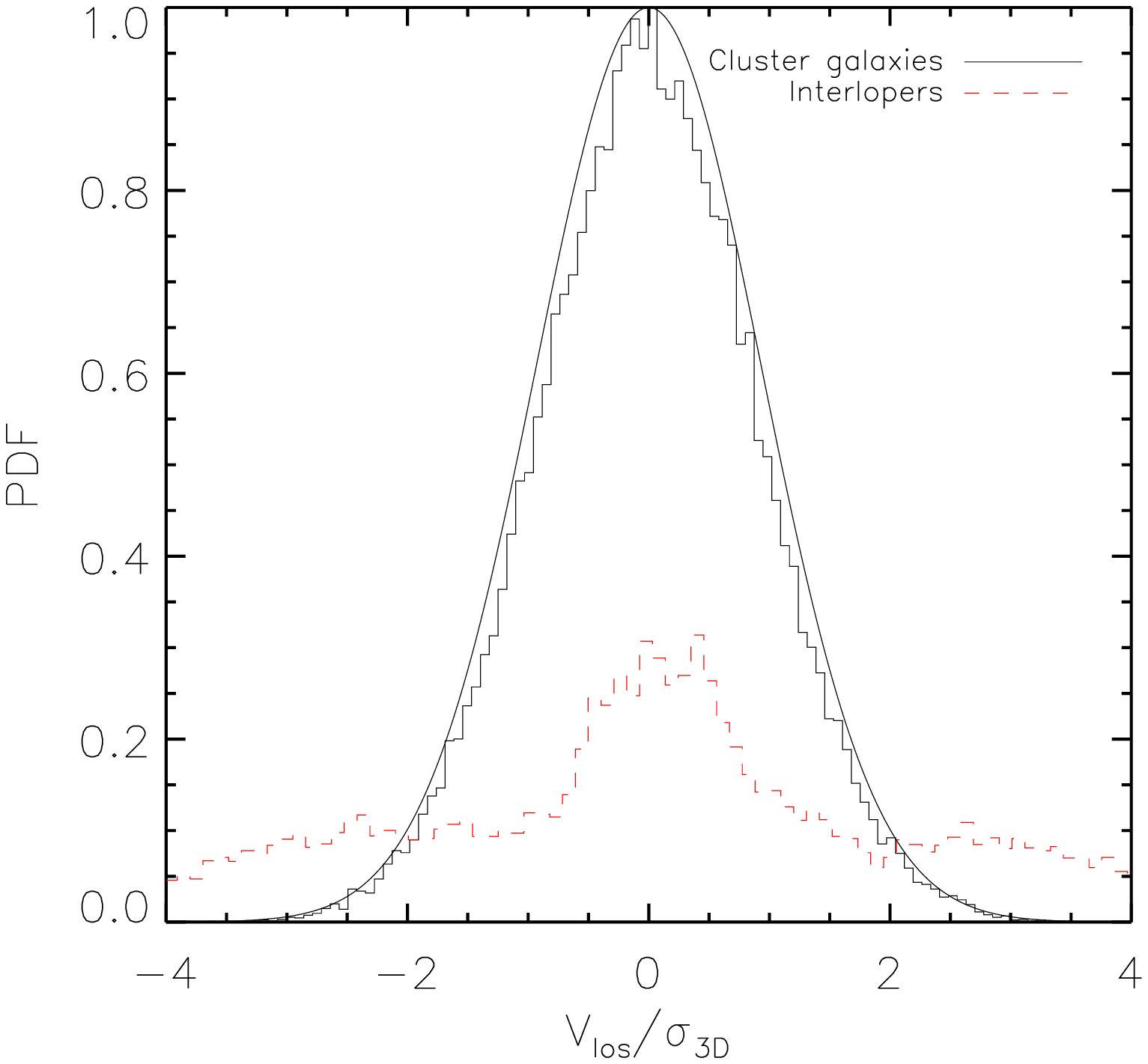,width=9.0cm}
    }
  \caption{{\it Left panel:} The color map represents the
    distribution of the line of sight velocity of cluster galaxies
    (within $3 \rvir$) normalized to the intrinsic 3D velocity
    dispersion of clusters as a function of the projected distance
    from the cluster center 
    in units of $\rvir$ for the sample described in the
    text. The contour lines represent the same
    distribution for the interloper galaxies. {\it Right panel:} The
    distribution of velocities in units of the intrinsic 3D velocity
    dispersion for the cluster galaxy population (solid black
    histogram) and for the interloper population (dashed red
    histogram). The normalisation is arbitrary, while the
    relative ratio of the two histograms reflects the sample described
    in the text. The solid black Gaussian is the expected distribution
    with width given by the Eq.~\ref{eq:meandyn} and $N_{gal} = 25$.}
  \label{fi:caustic}
\end{figure*}

To better understand how interlopers affect the inferred velocity
dispersion we select as an example all clusters with \mvir\ larger
than $5\times10^{14} \msun$. For each of the three orthogonal
projections we then initially select the most luminous 25 red-sequence
galaxies as described in Sect.~\ref{sec:selection} within a projected
distance of $1.5 \rvir$. We then apply the same procedure described
above to reject interlopers and obtain a final list of galaxies. From
this list of galaxies we then identify the "true" cluster members and
the interlopers. We show in the left panel of Fig.~\ref{fi:caustic} a
map representing the stacked distribution of the velocity of the
cluster galaxies as a function of the projected separation from the
cluster center $R_\perp/R_{vir}$. Note the typical trumpet shape
of the expected caustic distribution (\citealt{diaferio99},
\citealt{serra11}, \citealt{zhang11}). On the top of this map, we
overplot as contours the stacked distribution of the interloper
population that the $3\sigma$ clipping procedure was not able to
properly reject. A large fraction of high velocity interlopers are
still present after foreground and background removal and thus they
will bias high the estimated velocity dispersion.

This map highlights how caustic based techniques are potentially more
effective to remove interlopers than a simple $3\sigma$
clipping. However, observationally, a much larger number of galaxies
than the 25 spectra used here is typically needed to apply
these more sophisticated methods.

We also show in the right panel of Fig. \ref{fi:caustic} as solid
black and dashed red histograms respectively the distribution of
velocities for both the cluster galaxies and the interlopers
population. The expected Gaussian velocity distribution is overplotted
as a solid black Gaussian with a standard deviation given by
Eq. \ref{eq:meandyn} and $N_{gal} = 25$. The absolute normalisations of
the histograms are arbitrary, but the relative ratio of the two
histograms is representative of the ratio between the number of
cluster galaxies and interlopers. Note also that a
large fraction of low velocity interlopers is present. These
interlopers are mostly red-sequence galaxies which lie at about the
turn-around radius of the cluster over-density and therefore have
associated redshifts which are consistent with the cluster redshift.
As discussed above, a simple $3\sigma$ clipping technique is not able
to effectively remove high velocity interlopers, and therefore is
biasing high the inferred velocity dispersion. On the contrary caustic
based methods are able to remove this high velocity interlopers
population, but are not effective to reject this low velocity galaxies
at around the turn-around radius. As a net result, velocity
dispersions computed after interlopers rejection based upon caustic
techniques will be biased low (\citealt{wojtak07}, \citealt{zhang11}).

As mentioned above, for each cluster along all the projections we end
up with different samples of red-sequence galaxies that the $3\sigma$
clipping procedure recognises as "spectroscopic members".  Therefore,
for each different initially selected number of red-sequence galaxies,
we measure the robust estimation of the velocity dispersion. We then
apply Eq.~\ref{eq:fit} to estimate the dynamical mass. Left panel of
Fig.~\ref{fi:MAPS} shows the corresponding relation between the
resulting dynamical mass and the true virial mass for all the sample
stacked together. The dashed black-purple line is the one to one
relation, whereas the green lines show the 16, 50 and 84
percentiles. Note that the sample shown here is volume limited, and so
the distribution in mass is different than the typical observational
samples.  Furthermore, the same clusters appear several times with
dynamical masses computed from different number of galaxies on each
projection, and within different projected radii at all the
redshifts.  When red-sequence galaxies are selected within a
projected radius from a light-cone regardless of their true 3D
distance from the centre of the cluster, the relation between the
virial mass and the inferred dynamical mass is much broader.  In
particular, by looking at the median of the distribution, it is
possible to notice that a systematical overestimation of the dynamical
mass is present at all cluster masses, as expected from the interloper
contribution previously discussed. Furthermore, especially at the low
mass end of the cluster galaxy distribution, the presence of a
significant population of catastrophic outliers is making the relation
among virial mass and dynamical mass very asymmetric and causing a
severe boosting of the dynamical mass.

These outliers are likely related to cases where the simple $3\sigma$
clipping procedure is not sophisticated enough to effectively separate
the foreground and background interloper galaxies from the proper
cluster galaxies. To verify this hypothesis we show in the right panel
of Fig.~\ref{fi:MAPS} the same computation as for the left panel, but
restricting our sample to only the cases in which the presence of
interlopers is smaller than $5\%$.  We note how this sub-sample
qualitatively looks very similar to left panel of
Fig.~\ref{fi:fricran} which by construction contains only cluster
galaxies.  Furthermore, once the contribution from interlopers is
removed, the bias of dynamical mass over the true mass disappears.
However, remember that Fig.~\ref{fi:MAPS} shows that without
interlopers dynamical masses are on average underestimated compared to
the true virial mass, as expected from the effect of dynamical
friction described in Sect.~\ref{sec:dynamical}.  Moreover, as
expected from the lower panels of Fig.~\ref{fi:interlopers}, the
adopted interlopers rejection method is more effective for more
massive clusters.  Clearly the interloper effect on the dynamical mass
is more severe at the low mass end of the cluster population.

Because the color selection of cluster members is a crucial point
in this analysis, the results presented here obviously depend on the
adopted galaxy formation model at some level. On the one hand it is true
that the model is not perfectly reproducing the observed properties
of the cluster galaxy population. On the other hand we also do not
take into account any observational uncertainty which will instead
affect the real data, for example broadening the observed
red-sequence at fainter magnitudes.

To estimate the sensitivity of the color selection to uncertainties in
the galaxy modeling on the above described results, we select 
red-sequence galaxies with a different
criteria than the one described in Sect. \ref{sec:selection}.  Instead
of selecting the area in color-magnitude space which encompasses
68\% of the cluster galaxies, we select all galaxies within a fixed $\pm
0.15$ $mag$ along the fitted red-sequence relation, similarly to the
adopted criteria in the companion paper \citet{bazin12}. This is on
average a factor of $\sim 2$ in magnitude larger than the former
threshold (depending on the redshift ranging from $\sim 0.5 -3$, as
highlighted in Tab. \ref{t:cmr}). Then, we reject interlopers and
compute velocity dispersions and subsequent dynamical masses as
described in the above sections. We find that the fraction of
interlopers which the $3\sigma$ clipping procedure is not able to
reject is on average in agreement within $\sim 3\%$ with the previous
color selection. In particular for clusters with \mvir larger than
$4\times 10^{14} \mvir$ the agreement is better than 1\%.

We show in the left panel of Fig.~\ref{fi:MAPS} the resulting 16, 50
and 84 percentiles overplotted as red continuous lines. We note that a
larger effect from the interlopers is present in comparison with the
previous analyses, as expected from the broader color selection
adopted.  In particular, larger differences appear at the low mass end
of the cluster galaxy population, where a significant increase of
catastrophic outliers in the overestimation of the dynamical mass is
visible. On the other hand, the average population is not affected by
much. As a net result, a changing in the color selection of a factor
$\sim 2$ implies a change in the estimated velocity dispersion by less
than $\sim 3\%$. In particular, this difference reduces to less than $
\sim 1\%$ for clusters with \mvir larger than $5\times 10^{14}\msun$.

\subsection{Unbiased Dispersion Mass Estimator}
\label{sec:unbiased}

\begin{figure*}
  \centerline{\psfig{file=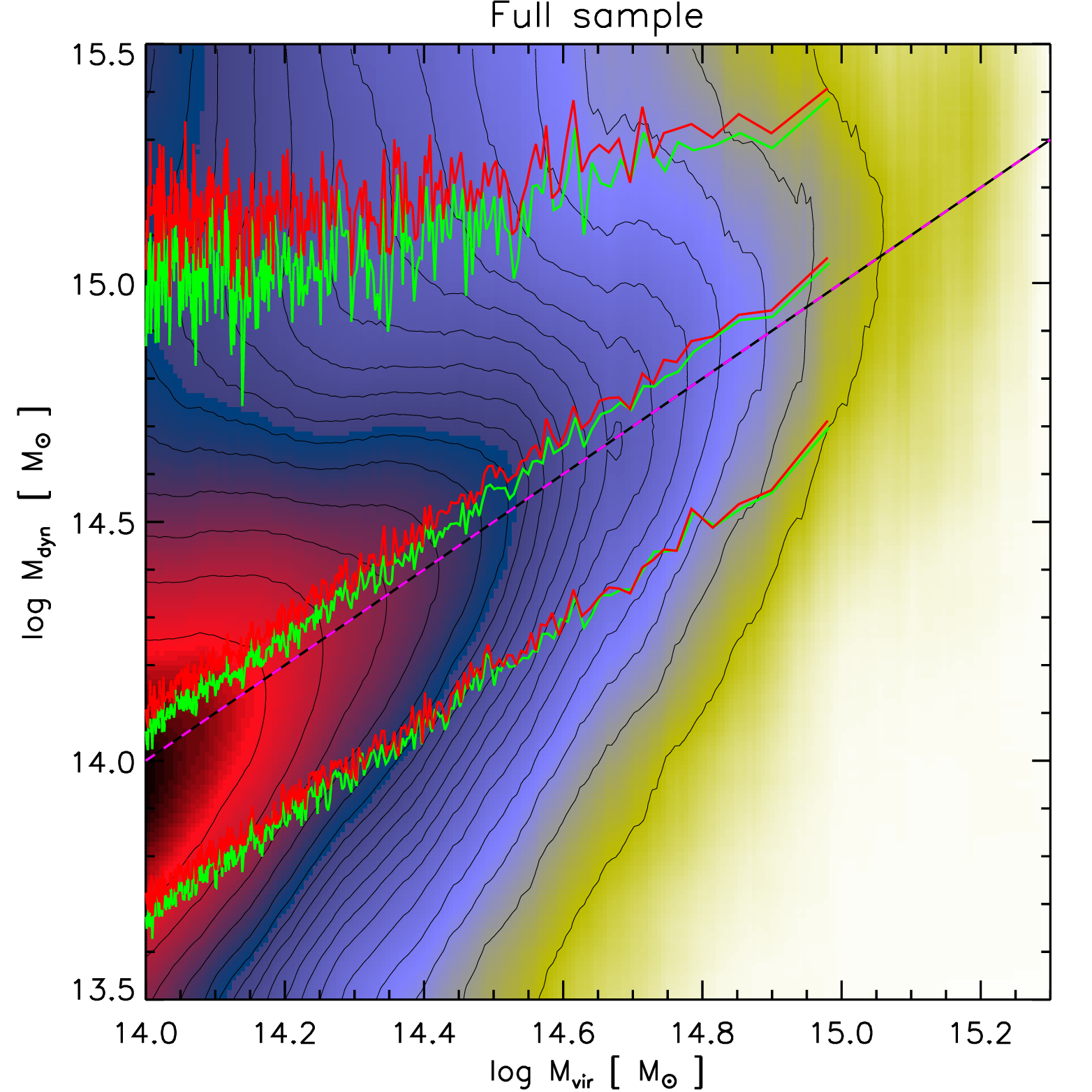,width=8cm}
      \psfig{file=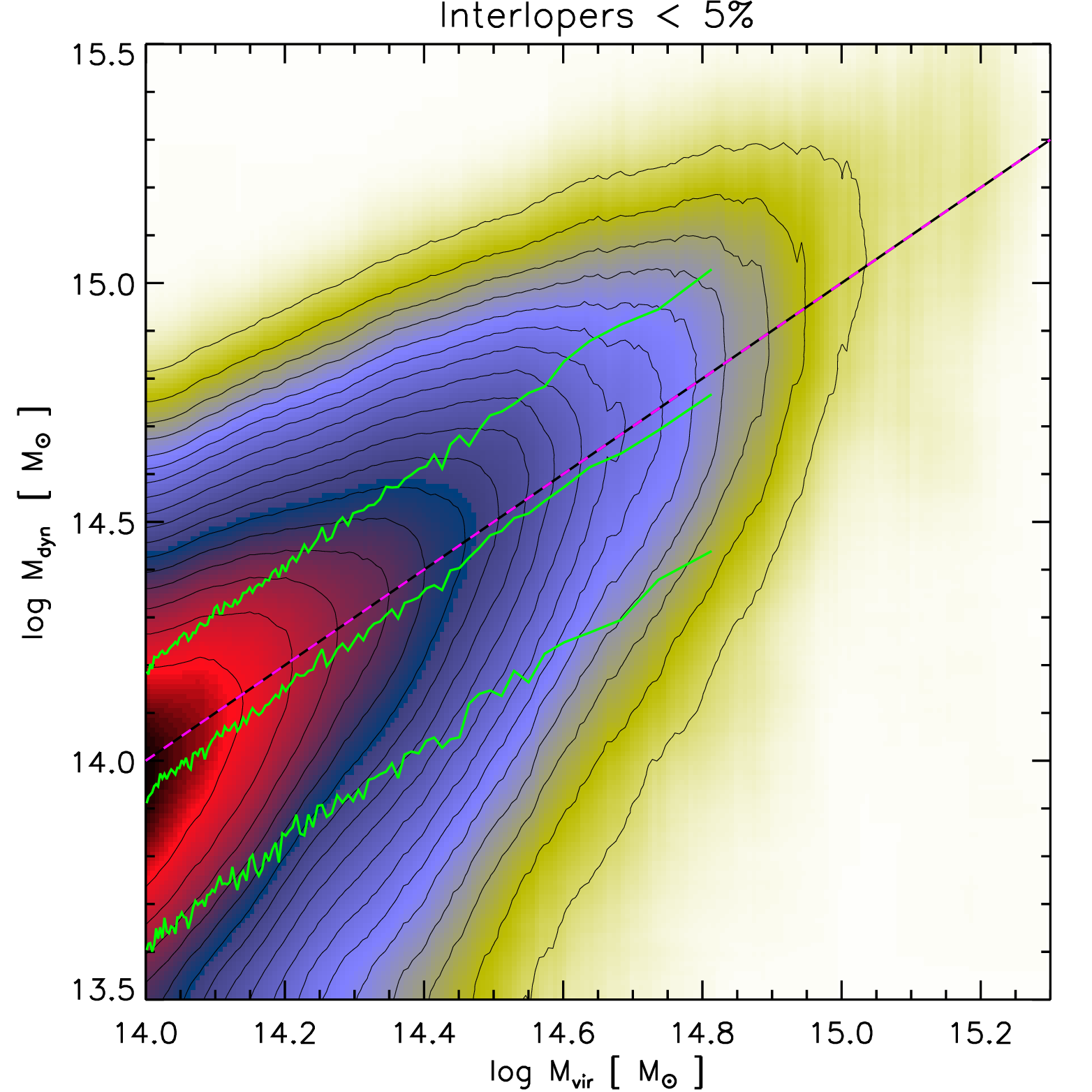,width=8cm}
   }
  \caption{{\it Left panel:} The distribution of the dynamical mass
    estimated through Eq. \ref{eq:fit} as a function of the true \mvir
    for the whole sample described in Sect. \ref{sec:selection} used
    in this work (green lines). Red lines represent the same
      distribution obtained from a different color selection of
      red-sequence galaxies as explained in Sect.
      \ref{sec:Interlopers}.{\it Right panel:} Same as for the left
    panel, but only for the cases where the fraction of interlopers is
    smaller than $5\%$. Dashed purple-black line is showing the
    one-to-one relation, while solid green and red lines are the 16,
    50 and 84 percentile.}
  \label{fi:MAPS}
\end{figure*}

\begin{figure}
  \hbox{
  \psfig{file=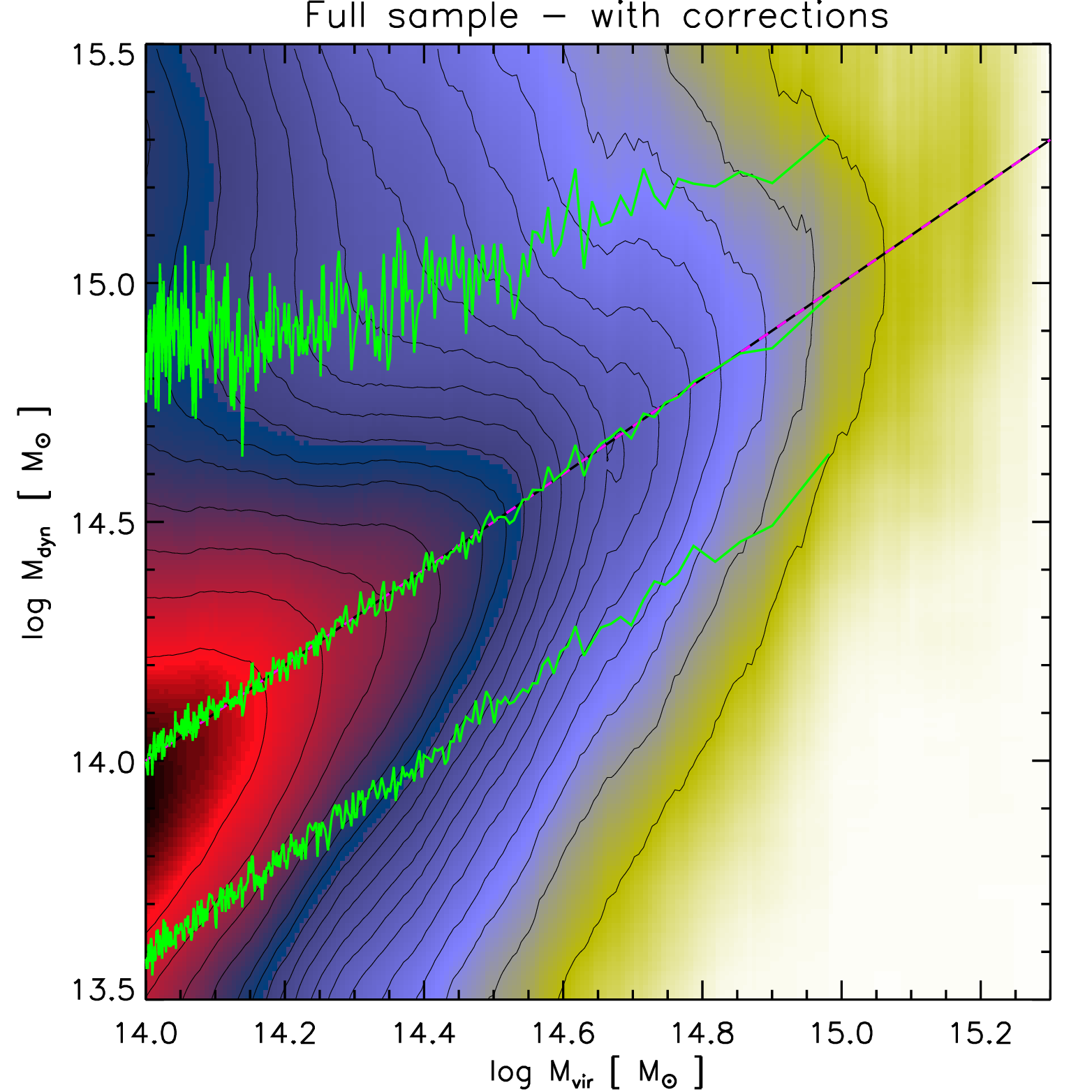,width=8cm}
  }
  \caption{Same as for the left panel of Fig.~\ref{fi:MAPS}, but
    velocity dispersions were numerically corrected as described in
    the text. Dashed purple-black line is showing the oat-to-one
    relation, while solid green lines are the 16, 50 and 84
    percentile.  Note that with these corrections the dynamical
    mass is an unbiased estimator of the true mass.}
  \label{fi:MAPSc}
\end{figure}

Similarly to Sect.~\ref{sec:dynamical}, we try to parametrise as a
function of the variables in Table \ref{tab:param} (aperture, number
of spectra, redshift and cluster mass), the way that interlopers
affect the inferred dynamical mass. However, we could not find a
satisfactory analytical solution to easily model the measured velocity
dispersion of clusters as a function of the above described variables,
due to the non-linear interplay of the explored parameter space
highlighted in Fig. \ref{fi:biasfrac}. Therefore, we numerically
compute the mean and the associated standard deviation of the ratio
between the observed and the 1D intrinsic velocity dispersion in
different bins of the parameter space as highlighted in Table
\ref{tab:param}. In this way, given the cluster mass, redshift and the
number of red-sequence galaxy spectra within a given projected radius
used to compute the velocity dispersion, we can correct for the
average bias affecting the estimation of the dynamical mass.  We show
in Fig. \ref{fi:MAPSc} the same relation described in the left panel
of Fig. \ref{fi:MAPS} when such corrections are included. We remark
that the bias is effectively removed at all the mass scales analysed
here. Furthermore, by comparing the 84 percentile and median lines at
the low mass end of the left panel of Fig.~\ref{fi:MAPS} with the ones
in Fig.~\ref{fi:MAPSc}, we note that while the former are separated by
about an order of magnitude in dynamical mass, for the later this
difference is reduced to about $0.8$ dex.

%%%%%%%%%%%%%%%%%%%%%%%%%%%%%%%%%%%%%%%%%%%%%%%%%%%%%%%%%%%%%%%%%%%%%%%%%%%%%%%

%%%%%%%%%%%%%%%%%%%%%%%%%%%%%%%%%%%%%%%%%%%%%%%%%%%%%%%%%%%%%%%%%%%%%%%%%%%%%%%
\section{Discussion and Conclusions}
\label{sec:Concl}

We have examined the use of velocity dispersions for unbiased mass 
estimation in galaxy clusters using the publicly available galaxy catalogue 
produced with the semi-analytic model by De Lucia \& Blaizot (2007) coupled with
the N-body cosmological Millennium Simulation (Springel et al. 2005).
In particular, we selected all galaxies in the SAM
with stellar mass larger than $10^8 \msun$ and analysed a sample
consisting of more than $\sim 20000$ galaxy clusters with $\mvir \geq
10^{14} \msun$ up to $z \sim 1.2$ (Tab: \ref{t:clus}).

First we explore the properties of the full galaxy sample and then we 
increase the level of complications to mimic the spectroscopic selection
that is typically undertaken in real world studies of clusters.  
Then we work through a series of controlled studies in an attempt 
to disentangle the different effects leading to biases and enhanced scatter
in velocity dispersion mass estimates.  Ultimately our goal is to inform the
dispersion based mass calibration of the SPT cluster sample (\citealt{bazin12}),
but we explore a broad range in selection in hopes that our results will
be of general use to the community.

Our primary conclusions for the full subhalo population are:

\begin{itemize}

\item We measure the galaxy (i.e. subhalo) velocity dispersion mass
  relation and show that it has low scatter ($\sim0.14$ in $ln(M)$)
  and that subhalo dispersions are $\lesssim$3\% lower than DM
  dispersion in \citet{evrard08}.  This difference corresponds to a $\lesssim
  10$\% bias in mass for our halos if the DM dispersion-- mass
 relation is used, and is consistent with previous determination
 of subhalo velocity bias.

\item We explore line of sight velocity dispersions of the full galaxy
  populations within the cluster ensemble and confirm that the
  triaxiality of the velocity dispersion ellipsoid is the dominant
  contributor to the characteristic $\sim$35\% scatter in dispersion
  based mass estimates.  We show that this scatter increases with
  redshift as $\sigma(z)\simeq 0.3+0.075z$.

\item We measure the principal axes and axial ratios of the spatial
  galaxy distribution ellipsoid, showing that there is a slight
  ($\sim5$\%) preference for prolate distributions; this property has
  no clear variation with mass or redshift.  We examine the line of
  sight velocity dispersions along the principle axes, showing that
  the slight preference toward prolate geometries translates into a
  slight ($\sim1$\%) bias in the dispersion mass estimates extracted
  from line of sight measures.

\end{itemize}

Our primary conclusions for the spectroscopic subsamples of subhalos are:

\begin{itemize}

\item We characterize the bias (Eqn.~\ref{eq:meandyn}) and the
  scatter (Eqn.~\ref{eq:sigmadyn}) in the line of sight velocity
  dispersion introduced by selecting a subset $N_{gal}$ of the most
  luminous red sequence galaxies within a cluster.  The bias is
  significant for samples with $N_{gal}<30$ and is likely due to
  dynamical friction of these most massive subhalos.  The scatter
  cannot be fully explained through a combination of intrinsic
  scatter in the relation between mass and the 3D dispersion of all
  galaxies (i.e. departures from equilibrium), scatter of the line of
  sight dispersion around the 3D dispersion (halo triaxility) and
  Poisson noise associated with the number of subhalos $N_{gal}$.
  A further component of scatter due to the presence of a dynamically cold 
  population of luminous red-sequence galaxies is needed to explain the
  full measured scatter.

\item We explore the impact of interlopers by creating spectroscopic
  samples using (1) red sequence color selection, (2) a maximum
  projected separation from the cluster center, and (3) $N$-sigma
  outlier rejection in line of sight velocity.  In these samples the
  interloper fraction (contamination) can be significant, growing from
  $\sim 10\%$ at the projected virial radius to $\sim$35\% at twice
  the project virial radius.  The contamination fraction has a much
  weaker dependency on the sample size $N_{gal}$.  We explore the
  dependence on mass and cluster redshift, showing that within a fixed
  aperture, contamination is a factor of $\sim 2$ worse at redshift
  $z\sim 1$ than at $z=0$.  Furthermore, we show that the fraction of
  interlopers is a steeper function of aperture for low mass clusters,
  but that at fixed redshift contamination does not change
  significantly with mass. We show that contamination is significant
  even if a more sophisticated caustic approach is used to reject
  interlopers, demonstrating that even clusters with large numbers of
  spectroscopic redshifts for red sequence selected galaxies suffer
  contamination from non-cluster galaxies.  We further study how
  interlopers are affecting the estimated velocity bias.  We find that
  the velocity bias has a minimum if computed within $ R_\perp \sim
  \rvir$. This is due to the balancing effect of larger intrinsic
  velocity bias at smaller radii and larger contamination at larger
  radii. Furthermore, we show that if velocity dispersions are
  computed within projected aperture $R_\perp$ larger than $\sim
  \rvir$, the velocity bias is a steeper function of $R_\perp$ for
  higher redshifts and lower cluster mass, as expected from the
  contamination fraction.

\item We study how changing the color selection affects the fraction
  of interlopers and the subsequent effect on the estimated velocity
  dispersion and dynamical masses.  We find that doubling the width of
  the color selection window centered on the red sequence has only a
  modest impact on the interloper fraction.  The primary effect of
  changing the color selection is on the filtering of catastrophic
  outliers.  This results in changes to the estimated velocity
  dispersion virial mass relation at the level of 1\% in mass.  We
  also show that uncertainties in the color selection are more
  important for low mass clusters than for the high mass end of the
  cluster population, which is because the dispersions of low mass
  clusters are more sensitive to catastrophic outliers.  The rather
  weak dependence of the dispersion based mass estimates on the
  details of the color selection suggests also that uncertainties in
  the star formation histories (and therefore colors) of galaxy
  populations in and around clusters are not an insurmountable
  challenge for developing unbiased cluster mass estimates from
  velocity dispersions.

\item We present a model to produce unbiased line of sight dispersion
  based mass estimates, correcting for interlopers and velocity bias.
  We also present the probability distribution function for the
  scatter of the mass estimates around the virial mass.  These two
  data products can be used together with a selection model describing
  real world cluster dispersion measurements to enable accurate
  cluster mass calibration.

\end{itemize}

In a companion paper, \citet{bazin12} apply this model in the
dispersion mass calibration of the SPT Sunyaev-Zel'odovich effect
selected cluster sample. We identify the following key remaining
challenges in using dispersions for precise and accurate mass
calibration of cluster cosmology samples. Surprisingly, the larger
systematic uncertainty has to be ascribed to our relative poor
knowledge of the velocity bias between galaxies or subhalos and DM. A
conservative estimate of this systematic is at the level of $< 5\%$
and arises from the comparison of different simulations and different
algorithms for subhalo identification (e.g. \citealt{evrard08}). The
systematic uncertainty in the color selection of galaxies and its
subsequent mapping between line of sight velocity dispersion and mass
is at relatively smaller level. Indeed we can estimate it at a $<1 \%$
level for samples selected as the ones described in \citet{bazin12},
despite the fact that galaxy formation models involve a range of
complex physical processes. In other words, systematics in predicting
galaxy properties (e.g. luminosity, colors, etc.) due to subgrid
physics associated with magnetic fields, AGN and supernova feedback,
radiative cooling and the details of star formation, do not appear to
significantly change the spectroscopic sample selection. On the other
hand, simulations including different physical treatments of gravity
are affecting the dynamics of the spectroscopic selected sample at a
higher level than we expected.  Given that the current dominant
contributor to the systematics floor is an issue associated with the
treatment of gravitational physics, there are reasons to be optimist
that future simulations will be able to reduce the current systematics
floor.

We acknowledge Jonathan Ruel for very useful discussions and support
from the Deutsche Forschungsgemeinschaft funded Excellence Cluster
Universe and the trans-regio program TR33: Dark Universe.

\label{lastpage}

\bibliographystyle{apj}
\bibliography{spt,paper}

\end{document}